\documentclass[prc,aps,superscriptaddress]{revtex4}  
\usepackage{graphicx} 
\usepackage{color}
\begin{document} 
\draft
\title{Momentum distributions and short-range correlations in the deuteron and $^3$He with
  modern chiral potentials}

\author{Laura Elisa Marcucci}
\affiliation{Dipartimento di Fisica ``Enrico Fermi'', Universit\`a
di Pisa, Largo Bruno Pontecorvo 3 - I-56127 Pisa, Italy}
\affiliation{Istituto Nazionale di Fisica Nucleare, Sezione di Pisa,\\
Largo Bruno Pontecorvo 3 - I-56127 Pisa, Italy}

\author{Francesca Sammarruca}
\affiliation{Department of Physics, University of Idaho, Moscow, ID 83844, USA}

\author{Michele Viviani}
\affiliation{Istituto Nazionale di Fisica Nucleare, Sezione di Pisa,\\
Largo Bruno Pontecorvo 3 - I-56127 Pisa, Italy}

\author{Ruprecht Machleidt}
\affiliation{Department of Physics, University of Idaho, Moscow, ID 83844, USA}
\date{\today} 
\begin{abstract}
  We study momentum distributions and short-range correlation
  probabilities
  in $A$=2 and $A$=3   
  systems.                                                                          
  First, we show results with phenomenological and meson-theoretic two- and 
  three-nucleon forces to verify consistency with previous similar studies.
  We then apply most recent high-quality chiral nucleon-nucleon
  potentials
  up to fifth
  order in the chiral expansion together with the leading chiral three-nucleon force.  
  Predictions are examined in the context of a broader discussion of
  short-range
  correlation
  probabilities extracted from analyses of inclusive electron scattering data,
  addressing the
  question of whether modern interactions can be reconciled with the latter. 
\end{abstract}
\maketitle 
        
\section{Introduction} 
\label{Intro} 

The study of high-momentum distributions in nuclei is fundamentally important
as it can reveal 
information about the short-range few-nucleon dynamics when the few-nucleon
system under consideration is surrounded by the medium. In this work, we focus on 
short-range correlation (SRC) in the deuteron and, with particular emphasis, in $^3$He. 

The lively discussion
recently stimulated by inclusive electron scattering measurements at high
momentum transfer on both light and heavy nuclei provides 
additional motivation for studying SRCs. In fact,                    
those measurements have been analyzed with the purpose of extracting
information on SRCs~\cite{CLAS,CLAS2,CLAS3,src,Pia+}.
In a suitable range of $Q^2$ (the four-momentum squared of the virtual photon) 
and of $x_B$ (the Bjorken variable), the cross section for the $A(e,e')X$
process is factorized such that cross section ratios for nuclei $A_1$
and $A_2$ can be related to the respective probability of a nucleon
to be involved in (either two-body or three-body) SRCs~\cite{Egi06}.
When extended to nuclear matter, this probability is equivalent to the
``wound integral'', which measures the amount of correlations in the
wave function and the $G$-matrix~\cite{FS14}. We recall, in passing, that the wound integral is the       
integral of the amplitude squared of the ``defect function", defined as the difference
between the correlated and the uncorrelated wave functions. 

Information about two-body SRCs can also be obtained in coincidence experiments
involving knock-out of a nucleon pair with protons~\cite{Tang} or
electrons~\cite{Korover,Shneor,Subedi,Baghda}. 

The plateaus seen in the ratios of inclusive scattering cross
section~\cite{CLAS,CLAS2} can be attributed 
to the dominance of SRCs for momenta above approximately 2 fm$^{-1}$. 
That is, when the electron scatters from a high-momentum nucleon
in the nucleus, the scattering can be viewed as an electron-deuteron
interaction, with the other $A-2$ nucleons essentially at rest.
More specifically, in an appropriate range of $Q^2$ and $x_B$, the ratio 
\begin{equation}
R(A_1,A_2) = \frac{\sigma(A_1,Q^2,x_B)/A_1}{\sigma(A_2,Q^2,x_B)/A_2} 
\label{xsec1}
\end{equation}
is expected to display scaling behavior. Under those circumstances, the cross section
ratio can be expressed as 
\begin{equation}
\frac{\sigma(A_1,Q^2,x_B)}{\sigma(A_2,Q^2,x_B)} = \frac{A_1}{A_2}R(A_1,A_2) \; , 
\label{xsec2}
\end{equation}
where $R$ is identified with the ratio               
of SRC probabilities in the two nuclei $A_1$ and $A_2$. Therefore,
measurements of inclusive electron 
scattering cross section ratios in the appropriate kinematical region can be related 
to the ratio of SRC probabilities, and ultimately
the absolute probability for a particular nucleus, given a suitable 
starting point, which, quite naturally, one would take to be the deuteron. 

Deuteron momentum distributions in the context of SRCs were studied in
Ref.~\cite{src2015} 
using local and non-local realistic two-nucleon (2N)
interactions. Those included:            
purely phenomenological local potentials, such as 
the Argonne $v_{18}$~\cite{av18} (AV18) or the Nijmegen II~\cite{Nij}
models, non-local meson-theoretic models, 
such as the charge dependent Bonn (CDBonn) potential~\cite{CD}, and state-of-the-art
non-local chiral potentials~\cite{EM03,chinn5,ME11}. In the study of Ref.~\cite{src2015},
it was concluded that predictions of high-momentum distributions in the
deuteron with non-local meson-exchange forces {\it or} state-of-the-art
chiral forces are systematically lower than those obtained
with the local AV18 or Nijmegen II potentials. Note that the AV18 predictions
were used in Refs.~\cite{CLAS,CLAS2}
to extract empirical information for heavier nuclei based on
Eqs.~(\ref{xsec1}) and~(\ref{xsec2}).

The analysis of Ref.~\cite{src2015} highlights 
non-localities in the tensor force as the source of differences in SRC among the various 
predictions, and suggests 
that such model dependence should be taken into account, as it may impact SRC considerations for heavier nuclei, 
see comments just below Eq.~(\ref{xsec2}).
At this point, it is appropriate to recall that the presence of non-locality 
in the tensor force has been found since a long time to be
a desirable feature       
in nuclear structure calculations. (For a discussion on the impact of 
non-locality in the one-pion exchange, see, for instance, Refs.~\cite{MSS96,Polls98,MP99}.) 

This paper contains updates and major extensions of the work of
Ref.~\cite{src2015}, 
presenting a simultaneous study of momentum distributions
in the deuteron and in $^3$He. First,             
we calculate the deuteron momentum distribution using the most recent 
chiral 2N potentials from Ref.~\cite{EMN},
from leading to fifth chiral order.
These interactions are better and more consistent than the
ones of Refs.~\cite{EM03,chinn5,ME11} used in
Ref.~\cite{src2015}, because the same power counting scheme and
cutoff procedures are used at each order. In addition, the $\pi N$
low-energy constants (LECs) are                 
the very accurate ones determined in the Roy-Steiner analysis of
Ref.~\cite{Hofe+}. The uncertainty associated with these LECs
is sufficiently small that variations within their errors have negligible 
impact on the construction of the potentials, which 
are non-local and of soft nature. A point worth mentioning is that
these 2N forces can predict a triton binding energy around 8.1 MeV,
leaving only very small room for three-nucleon (3N) forces.  

We then proceed to consider the single-nucleon (1N)
and 2N momentum distributions in $^3$He using
the phenomenological AV18 and the meson-theoretic CDBonn potentials, 
alone or augmented by 3N forces, namely the Urbana IX (UIX)
model~\cite{UIX}
in conjunction with AV18, and
the Tucson-Merlbourne
(TM)~\cite{TM} 3N force in conjunction with CDBonn.
This will allow us to quantify the 3N force contributions within 
the framework of these older forces. To verify our calculations, 
results obtained with the AV18 and AV18/UIX potential models will be compared
with the previous studies of Refs.~\cite{Alv13,Alv16,Wir_web}.

Having established a reliable baseline, 
we shift our focus to the more novel aspects of this 
work, namely the most recent high precision chiral 2N
potentials~\cite{EMN} and corresponding chiral 3N force.
The main motivation behind this calculation can be explained
as follows.
The presence of high-momentum components in the nuclear wave function
is an indication of              
SRCs.  At the two-body level, SRCs originate from the (repulsive)
short-range central and tensor force, which,       
in the well-established and still popular meson-exchange phenomenology,
are described by $\omega$- and $\rho$-meson exchange, respectively.        
Although 
realistic meson-theoretic or purely phenomenological interactions               
are frequently employed in contemporary calculations 
of nuclear structure and reactions, this approach has some intrinsic
problems/limitations.
First, the connection between the 2N and the applied 3N force does not rest
on firm grounds.
Second, no clear mechanism exists to quantify and control the theoretical
uncertainty of a prediction. 
These problems are absent from the 
chiral effective field theory ($\chi$EFT)
approach, which provides a well-defined prescription to develop nuclear forces 
in an internally consistent manner at each order of a systematic perturbative
expansion.
In fact, using effective degrees of freedom, namely
hadrons (nucleons and pions), and
maintaining a link with quantum chromodynamics (QCD)
through the symmetries of low-energy QCD, $\chi$EFT has become a
well-established and, in principle, 
model-independent framework to develop nuclear forces and
quantify the theoretical uncertainty at each order of the expansion. 
Therefore, we find it both important and insightful to perform these     
calculations using state-of-the-art chiral interactions. 

The paper is organized as follows: In Sec.~\ref{sec:II} we set the stage with a brief discussion on     
the deuteron, while we address $^3$He
in Sec.~\ref{sec:III}. In the latter section,
we will first present a brief review of the numerical
techniques used to calculate the $A=3$ nuclear wave functions
and the 1N and 2N momentum distributions. Then
we will show and discuss results obtained with the older                  
AV18 and CDBonn potential models, augmented or not by the UIX~\cite{UIX}
and the TM~\cite{TM}
3N force, respectively, as well as the chiral 2N potentials
of Ref.~\cite{EMN}, without or with the chiral 3N force.
We will also discuss the procedure adopted to
determine the two LECs entering the leading 3N force.                    
Our conclusions and future plans are summarized in Sec.~\ref{sec:concl}.
     
\section{High-momentum distribution and SRCs in the deuteron} 
\label{sec:II} 

\begin{figure}[!t] 
\centering         
\vspace*{1.2cm}
\hspace*{0.5cm}
\scalebox{0.35}{\includegraphics{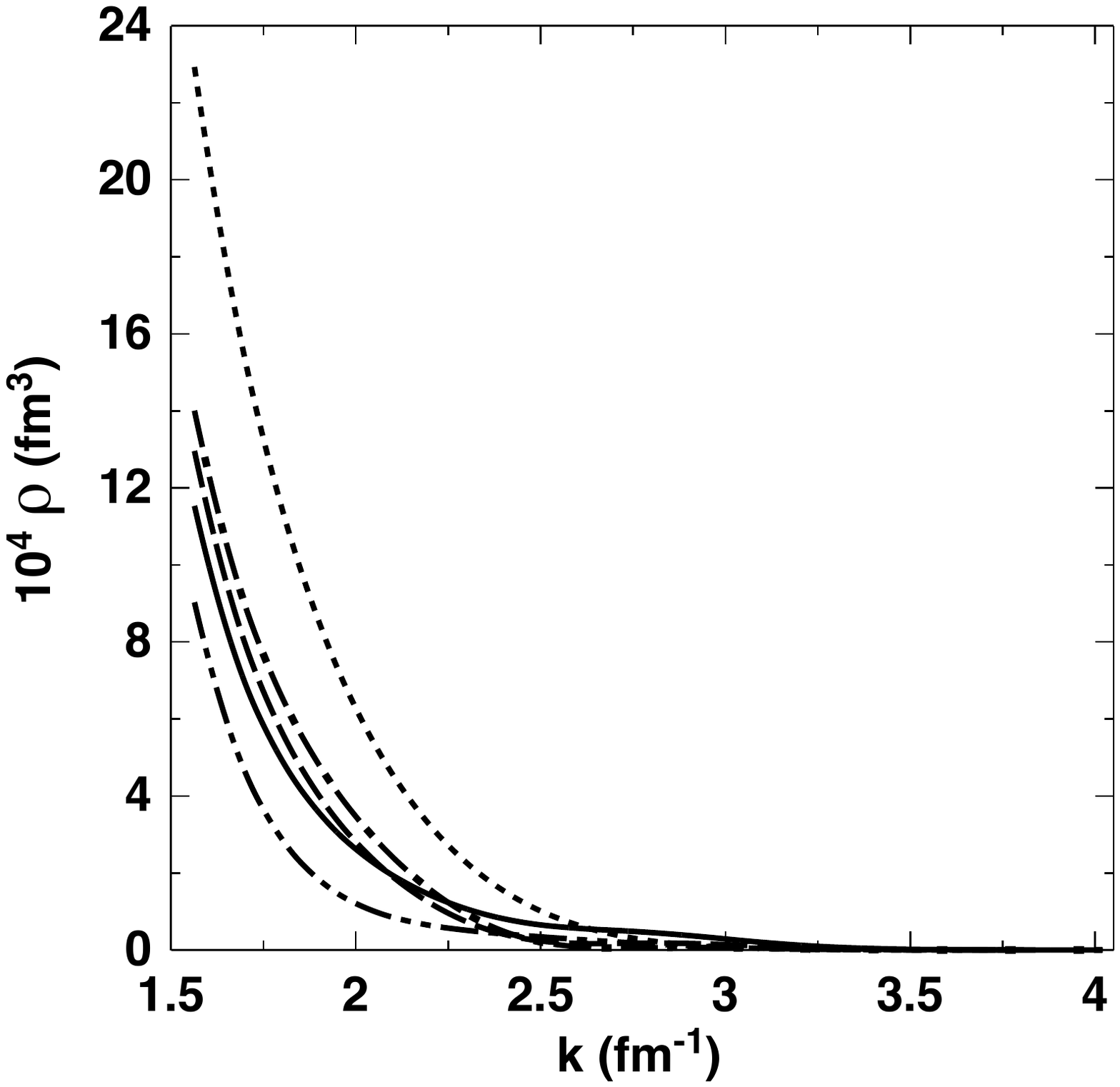}}
\scalebox{0.35}{\includegraphics{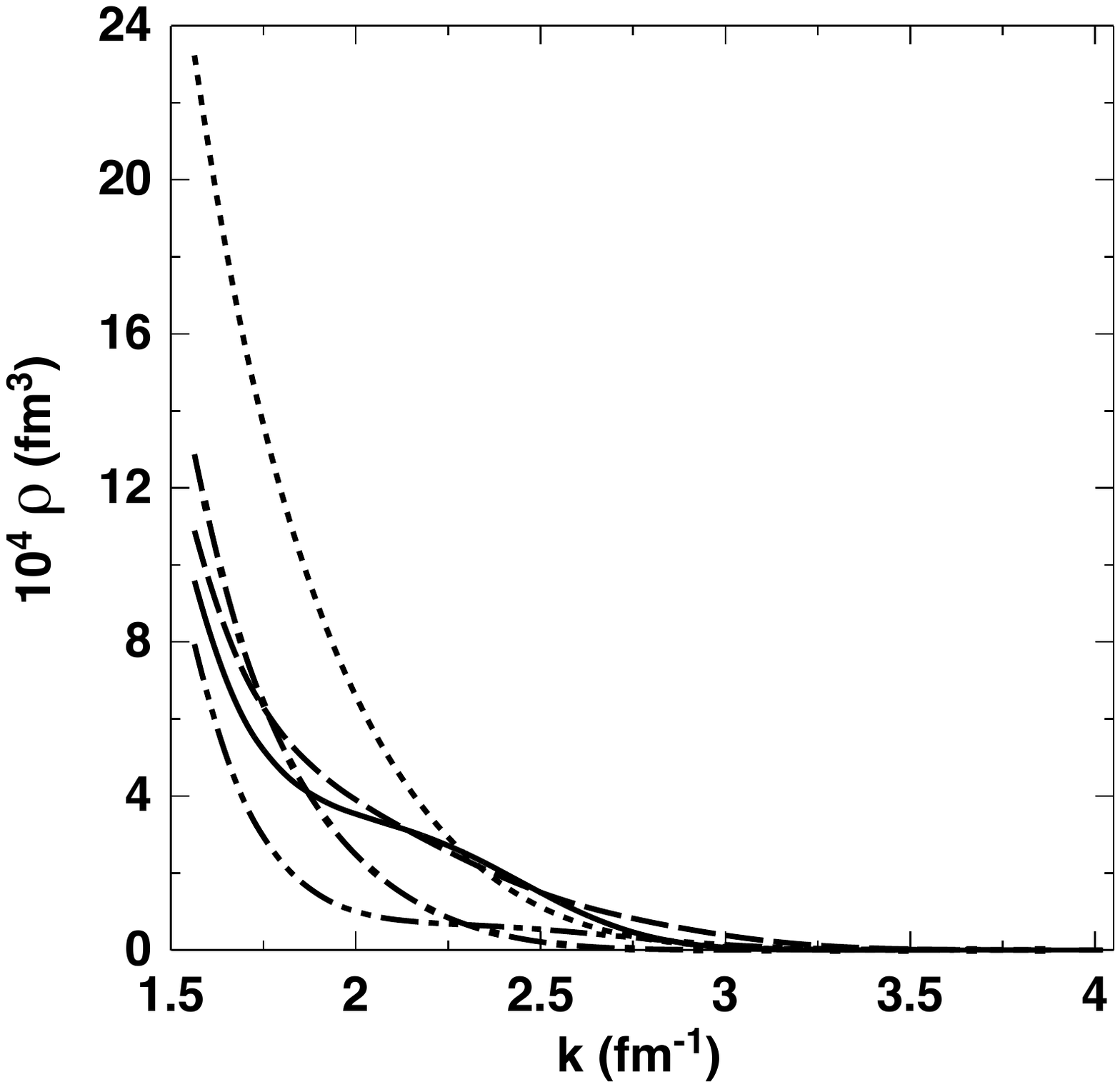}}
\vspace*{-3.0cm}
\caption{Left panel: Momentum distributions in the
  deuteron predicted with the chiral potentials 
  of Ref.~\cite{EMN} at LO (dotted), NLO (dash-double dot),
  N$^2$LO (dash-dot), N$^3$LO (dash), N$^4$LO (solid).
  The cutoff is fixed at $\Lambda=500$ MeV.
  Right panel: Predictions taken from Ref.~\cite{src2015}, and are obtained
  using the chiral potentials of Ref.~\cite{EM03,chinn5,ME11}.
} 
\label{FIG1}
\end{figure}

\begin{table}                
  \centering
  \caption{Probabilities of SRCs as defined
    in Eq.~(\ref{psrc}) and deuteron $D$-state percentage
    for the chiral interactions considered  on
    the left panel of Fig.~\ref{FIG1}.         
    The values in parenthesis, given
    for comparison, are taken from Ref.~\cite{src2015} and
    correspond to the distributions shown on the right 
    panel of Fig.~\ref{FIG1}. The cutoff $\Lambda$ is equal to 500 MeV in all cases.} 
\begin{tabular}{c|c|c}
\hline\hline
  Model & $a_{2N}(d)$ & $P_D$ \\
\hline\hline
 LO &  0.046 (0.047)& 0.0729 (0.0757)  \\
    &   &  \\ 
 NLO &  0.015 (0.015)& 0.0340 (0.0313)  \\
   &    &   \\ 
 N$^2$LO &  0.026 (0.022)& 0.0449 (0.0417)  \\
   &    &    \\ 
 N$^3$LO &  0.024 (0.030) & 0.0415 (0.0451)  \\
  &    &    \\ 
 N$^4$LO &  0.024 (0.026)& 0.0410 (0.0414) \\
\hline\hline
\end{tabular}
\label{tab1} 
\end{table}

To best put this study in context, 
we begin with a quick review of the 2N bound state.                     
In particular, we present in Fig.~\ref{FIG1} the deuteron momentum
distributions $\rho(k)$,
defined as the Fourier transform squared of the
coordinate-space deuteron wave function.                    
On the left side of the figure, we show the results,
with focus on high-momentum components, obtained with the
latest chiral interactions of Ref.~\cite{EMN}                  
from leading to fifth order (N$^4$LO).
On the right side of the figure, we show for comparison the same quantities
calculated as in Ref.~\cite{src2015} with the
older chiral potentials of Refs.~\cite{EM03,chinn5,ME11}.
From inspection of the figure,
we can conclude that the convergence pattern
has definitely improved with the new potentials. 

We then define the probability of SRCs in the deuteron as in
Ref.~\cite{src2015}, i.e.\
\begin{equation}
a_{2N}(d) = 4 \pi \int _{k^-}^{\infty} \rho(k)k^2\;dk \; , 
\label{psrc} 
\end{equation} 
where                                                                        
$k^-$ is taken to be 1.4 fm$^{-1}$ (276 MeV). This definition was
adopted in Ref.~\cite{CLAS}, where the choice of the lower integration limit
was suggested by the onset of scaling of the cross section, which
should signal the dominance of scattering from a strongly correlated nucleon.
In view of Eqs.~(\ref{xsec1})--(\ref{xsec2}), 
the absolute per-nucleon SRC probability in a nucleus $A$ can be deduced
if the absolute per-nucleon probability in $^3$He and the deuteron
are calculated or estimated. More precisely,    
\begin{equation}
  a_{2N}(A)=a_{2}(A/^3{\rm He})a_{2N}(^3{\rm He})
  \; \; \; \; \mbox{and} \; \; \; \;
  a_{2N}(^3{\rm He})=a_{2}(^3{\rm He}/d)a_{2N}(d) \;, 
  \label{eq:a2nA}
\end{equation}
where $a_2(A_1/A_2)$ is the SRC probability for nucleus $A_1$
relative to nucleus $A_2$. 
The probability in the deuteron was taken to be equal to
$0.041\pm 0.008$ in Ref.~\cite{CLAS2}.
We list in Table~\ref{tab1} the integrated probabilities $a_{2N}(d)$
defined in Eq.~(\ref{psrc}),
calculated integrating the curves of Fig.~\ref{FIG1} (left panel). 
As an additional, related information, we also show the corresponding
$D$-state percentage.
In fact, deuteron $D$-state probabilities are larger with stronger           
short-range central and tensor components of the nuclear
force which, for the non-local
chiral interactions and, generally, for non-local interactions, are softer
than for the local AV18 potential.
The values in parenthesis correspond to the distributions displayed on the
right of Fig.~\ref{FIG1}, i.e.\ obtained with the older chiral
potentials of Refs.~\cite{EM03,chinn5,ME11}.
As the table shows, there are huge variations
between the LO and the NLO cases, and still large differences between NLO and
N$^2$LO.
Variations at higher orders indicate                                   
a clear convergence pattern, definitely
improved by the use of the newest potentials.            
Finally we notice that the deuteron integrated probabilities
$a_{2N}(d)$ 
display significant model-dependence, as the
corresponding values obtained with the AV18 and the CDBonn potentials
are 0.042 and 0.032, respectively.                                  
We will show below that similar considerations apply to $^3$He as well. 
This model dependence is likely to propagate in the analyses for
heavier nuclei.

\section{High-momentum distribution and SRCs in the $^3$He nucleus} 
\label{sec:III}

\subsection{Theoretical formalism}
\label{subsec:th}

We briefly review the method used to solve the
$A=3$ quantum mechanical problem, i.e. the Hyperspherical
Harmonics (HH) method. This method has the great advantage that
we can work both in coordinate- and momentum-space, with no restriction
on the choice of the nuclear potential model, either local or
non-local. The starting point are the so-called Jacobi coordinates,
which are defined in coordinate-space as~\cite{Viv06,Kie08}
\begin{eqnarray}
  {\bf x}_p &=& \frac{1}{\sqrt{2}} ({\bf{r}}_j - {\bf{r}}_i) \ ,
  \nonumber \\
  {\bf y}_p &=& \sqrt{\frac{2}{3}} \left({\bf{r}}_k-\frac{1}{2}
            ({\bf{r}}_i + {\bf{r}}_j)\right) \ ,
    \label{eq:jacobir}
\end{eqnarray}
where $p$ represents an even permutation of $i,j,k=1,2,3$, with
$p=1$ for $i,j,k=2,3,1$,
and
${\bf r}_i$ is the position of the $i$-th particle.
The conjugate Jacobi momenta (in unit of $\hbar=1$) are defined as 
\begin{eqnarray}
  {\bf q}_p &=& \frac{1}{\sqrt{2}} ({\bf{p}}_j - {\bf{p}}_i) \ ,
  \nonumber \\
  {\bf k}_p &=& \sqrt{\frac{2}{3}} \left({\bf{p}}_k-\frac{1}{2}
            ({\bf{p}}_i + {\bf{p}}_j)\right) \ ,
  \label{eq:jacobip}
\end{eqnarray}
${\bf{p}}_i$ being the momentum of the $i$-th particle.
The next step is to introduce the so-called hyperradius $\rho$
and hypermomentum $Q$ as
\begin{eqnarray}
  \rho&=&\sqrt{{\bf x}_p^2+{\bf y}_p^2}\ , \nonumber \\
  Q&=&\sqrt{{\bf k}_p^2+{\bf q}_p^2}\ ,
  \label{eq:hyperrp}
\end{eqnarray}
and the hyperangle $\phi_p^{(\rho /Q)}$, given by
\begin{eqnarray}
  \tan\phi_p^{(\rho)}&= & \frac{y_p}{x_p} \ , \nonumber \\
  \tan\phi_p^{(Q)}&= & \frac{k_p}{q_p} \ .  
  \label{eq:hyperphi}
\end{eqnarray}
We note that $\rho$ and $Q$ do not depend on the considered
permutation, while $\phi_p^{(\rho)}$ or $\phi_p^{(Q)}$ do.
Then, the HH functions for the $A=3$ system are given in coordinate
space by
\begin{equation}
  {\cal Y}_{\alpha,n}(\Omega^{(\rho)}_p)=[[Y_l({\hat{\bf x}}_p)\otimes
    Y_L({\hat{\bf y}}_p)]_{\Lambda \Lambda_z}\otimes
    [\chi_{S_{ij}}\otimes \frac{1}{2}]_{\Sigma\Sigma_z}]_{JJ_z}
  [\eta_{T_{ij}}\otimes \frac{1}{2}]_{TT_z}
  P_{n,l,L}(\phi_p^{(\rho)})
    \ ,
  \label{eq:HH}
\end{equation}
where $\Omega^{(\rho)}_p=(\phi_p^{(\rho)},{\hat{\bf x}}_p,{\hat{\bf y}}_p)$
and
\begin{equation}
  P_{n,l,L}(\phi_p^{(\rho)})=N_{n,l,L}(\cos\phi_p^{(\rho)})^l (\sin\phi_p^{(\rho)})^L
  P_n^{L+1/2,l+1/2}(\cos 2\phi_p^{(\rho)})\label{eq:Jacobipol} \ ,
\end{equation}
$N_{n,l,L}$ being a normalization coefficient and
$P_n^{L+1/2,l+1/2}(\cos 2\phi_p^{(\rho)})$ a Jacobi polynomial of degree $n$.
In Eq.~(\ref{eq:HH}), $Y_l({\hat{\bf x}}_p)$ and $Y_L({\hat{\bf y}}_p)$
are spherical harmonics in the two Jacobi coordinates,
coupled to the total orbital angular momentum $\Lambda,\Lambda_z$, 
$\chi_{S_{ij}}$
($\eta_{T_{ij}}$) is the spin (isospin) function
of the pair $ij$, where the spins (isospins) of the particles
$i$ and $j$ are coupled to $S_{ij}$ ($T_{ij}$), which is itself coupled
to the spin (isospin) 1/2 of particle $k$ to give the total
spin (isospin) $\Sigma,\Sigma_z$ ($T,T_z$). The total orbital angular
momentum $\Lambda$ and the total spin $\Sigma$ are coupled
to the total angular momentum $J,J_z$. Finally,
we remark that the grand-angular momentum is defined
as $G=2n+l+L$, and we have labelled with the channel index
$\alpha$ the set of quantum numbers $[l,L,\Lambda,S_{ij},\Sigma,T_{ij},T]$
which determine the spin-isospin-angular state.
An expression similar to Eq.~(\ref{eq:HH}) holds in momentum-space,
with appropriate changes.

Having introduced the HH functions, the $A=3$ nuclear
wave function can be written as
\begin{equation}
  \Psi=\sum_{\alpha,n}u_{\alpha,n}(\rho)\sum_p{\cal{Y}}_{\alpha,n}(\Omega_p^{(\rho)})
  \label{eq:psir} \ ,
\end{equation}
where $u_{\alpha,n}(\rho)$ is the hyperradial function to be determined.
Similarly, in momentum-space we can write
\begin{equation}
  \Psi=\sum_{\alpha,n}w_{\alpha,n}(Q)\sum_p{\cal{Y}}_{\alpha,n}(\Omega_p^{(Q)})
  \label{eq:psip} \ ,
\end{equation}
where $w_{\alpha,n}(Q)$ is the function of the hypermomentum $Q$,
and it is related to $u_{\alpha,n}(\rho)$ via essentially a Fourier
transform~\cite{Viv06}, i.e.\
\begin{equation}
  w_{\alpha,n}(Q)=(-i)^G\int_0^\infty d\rho\,\frac{\rho^5}{{Q\rho}^{2}}\,
  {\cal J}_{G+2}(Q\rho) u_{\alpha,n}(\rho) \label{eq:wu} \ ,
\end{equation}
where ${\cal J}_{G+2}(Q\rho)$ is a Bessel function.
Finally, the functions $u_{\alpha,n}(\rho)$ (or $w_{\alpha,n}(Q)$) are
themselves expanded on a basis of Laguerre polynomials (or their Fourier
transform) as
\begin{eqnarray}
  u_{\alpha,n}(\rho)&=&\sum_k c_{\alpha,n,k}\,\,\, ^{(5)}L_k(\gamma\rho)\,
  {\rm e}^{-\gamma\rho/2} \ , \label{eq:urho}
\end{eqnarray}
where $c_{\alpha,n,k}$ are unknown coefficients and
$\gamma$ is a non-linear parameter, chosen to
be 4 fm$^{-1}$ for the local AV18 or AV18/UIX potentials,
and 7 fm$^{-1}$ for the other non-local potentials.
These values are the ones used in Refs.~\cite{Viv06,Kie08}.
Equations~(\ref{eq:psir})--(\ref{eq:urho}) can be cast in a compact form as
\begin{equation}
  \Psi=\sum_\mu c_\mu \phi_\mu \label{eq:psi2} \ ,
\end{equation}
where $\phi_\mu$ are given either in coordinate- or momentum-space.
What is essential is that the $c_\mu$ coefficients of the expansion
are the same in both cases. These coefficients
are determined using the Rayleigh-Ritz
variational principle, and the problem of determining $c_\mu$ and the
energy $E$ is reduced to a generalized eigenvalue problem,
\begin{equation}
  \sum_{\mu^\prime}c_{\mu^\prime}\langle \phi_\mu | H-E | \phi_{\mu^\prime}\rangle = 0
  \ .
  \label{eq:rrvar}
\end{equation}
The advantage of having $\phi_\mu$ expressed either in coordinate- or
in momentum-space is clear: the matrix elements of local operators
will be calculated in coordinate-space, those
of non-local operators in momentum-space. 
Furthermore, the
1N and 2N momentum-distributions can be written straightforward in
momentum-space,
without the need to perform any additional Fourier transform,
unlike what was done in Refs.~\cite{Alv13,Alv16,Wir_web}. We will define and    
evaluate these momentum-distributions in the next sections.

We conclude this section by discussing the construction of the 3N force
in the chiral approach. As is well known,
the chiral 3N force appears for the first time at N2LO.
It consists of three contributions:
the two-pion exchange (2PE) term, the one-pion exchange (1PE)
diagram, and a short-range contact term. The 1PE and the contact terms
are multiplied by two LECs, $c_D$ and $c_E$, respectively. We
determine them within a well established procedure (see
Ref.~\cite{Mar12} and references therein), repeated in
Ref.~\cite{nuclmatt18} for the new chiral
potentials of Ref.~\cite{EMN}. In particular, the
LECs  $c_D$ and $c_E$ are constrained to reproduce the
$A=3$ binding energies and the Gamow-Teller (GT) matrix element of
tritium $\beta$-decay. For completeness,
the values of $c_D$ and $c_E$ from Table I and II
of Ref.~\cite{nuclmatt18} are reported again here in
Table~\ref{tab:cdceI} and~\ref{tab:cdceII}, which include, in addition, 
the values obtained
with $\Lambda=550$ MeV.                       
In the first table, the $c_D$ and $c_E$ values are obtained
using the 3N force up to N2LO.
The complete 3N force beyond N2LO is very complex and
often neglected in nuclear structure studies. However,
the 2PE component of the 3N force can be                
calculated fully up to N4LO. In Ref.~\cite{Kre12} it was shown that
the 2PE 3N force has essentially the same analytical structure
at N2LO, N3LO, and N4LO. Thus, one can add the
three orders of this 3N force component and parametrize the
result in terms of effective LECs. These effective LECs
are taken from Table IX of Ref.~\cite{EMN}
and shown here in Table~\ref{tab:cdceII}. 
By using these $c_{1,3,4}$ in the mathematical
expression of the N2LO 3N force, one can include the 2PE
parts of the 3N force up to N3LO and N4LO in a simple way.
Obviously, the LECs $c_D$ and $c_E$ are fitted
again for each case and are listed in Table~\ref{tab:cdceII}.
The error arising from the
fitting procedure, shown 
in parentheses, is quite large. On the other hand, we have observed                           
that the impact of the 3N interaction on the           
momentum distributions and SRCs is weak (see below). Thus, we find it
appropriate to
use in our study the wave functions obtained
adopting the central values of $c_D$ and $c_E$.
\begin{table}[t]
\centering
\caption{Values for the LECs $c_{1,3,4}$, $c_D$ and $c_E$ at the chiral
  orders N2LO, N3LO and N4LO. The $c_D$ and $c_E$ LECs
  reproduce the $A=3$ binding energies and the GT
  matrix element in tritium $\beta$-decay, as explained
  in the text. The 2PE N3LO 3N interactions are
  not included, i.e.\ the $c_{1,3,4}$ LECs in the 3N force are those
  of Table II of Ref.~\cite{EMN}.
  The numbers in
  parentheses indicate the error arising from the fitting procedure.}
\begin{tabular}{ccccccc}
\hline
\hline
  & $\Lambda$ (MeV) & $c_1$ & $c_3$ & $c_4$ & $c_D$ & $c_E$ \\
\hline    
\hline
N2LO & 450 &  --0.74 & --3.61 & 2.44  &    0.935(0.215) &   0.12(0.04)  \\
     & 500 &  --0.74 & --3.61 & 2.44  &    0.495(0.195) & --0.07(0.04) \\
     & 550 &  --0.74 & --3.61 & 2.44  &  --0.140(0.190) & --0.44(0.03) \\
\hline 
N3LO & 450 &  --1.07 & --5.32 & 3.56  &    0.675(0.205) &   0.31(0.05)  \\
     & 500 &  --1.07 & --5.32 & 3.56  &  --0.945(0.215) & --0.68(0.04) \\
     & 550 &  --1.07 & --5.32 & 3.56  &  --1.610(0.220) & --1.69(0.03) \\
\hline 
N4LO & 450 &  --1.10 & --5.54 & 4.17  &  1.245(0.225) &   0.28(0.05)  \\
     & 500 &  --1.10 & --5.54 & 4.17  &--0.670(0.230) & --0.83(0.03)  \\
     & 550 &  --1.10 & --5.54 & 4.17  &--1.245(0.175) & --1.91(0.02)  \\
\hline
\hline
\end{tabular}
\label{tab:cdceI}
\end{table}

\begin{table}
  \caption{Same as Table~\ref{tab:cdceI} but including
    the 2PE 3N interaction at N3LO and N4LO,
    i.e.\ the $c_{1,3,4}$ LECs in the 3N force are those
    of Table IX of Ref.~\cite{EMN}.}
    \label{tab:cdceII}
\begin{tabular}{ccccccc}
\hline
\hline
   & $\Lambda$ (MeV) & $c_1$ & $c_3$ & $c_4$ & $c_D$ & $c_E$ \\
\hline     
\hline
N2LO & 450 &  --0.74 & --3.61 & 2.44 &  0.935(0.215) &  0.12(0.04) \\
     & 500 &  --0.74 & --3.61 & 2.44 &  0.495(0.195) &--0.07(0.04) \\
     & 550 &  --0.74 & --3.61 & 2.44  &  --0.140(0.190) & --0.44(0.03) \\
\hline 
N3LO & 450 &  --1.20 & --4.43 & 2.67 &  0.670(0.210) &  0.41(0.05) \\
     & 500  & --1.20 & --4.43 & 2.67 &--0.750(0.210) &--0.41(0.04) \\
     & 550  & --1.20 & --4.43 & 2.67 &--1.350(0.200) &--1.14(0.03) \\
\hline
N4LO & 450 &  --0.73 & --3.38 & 1.69 &  0.560(0.220) &  0.46(0.05) \\
     & 500 &  --0.73 & --3.38 & 1.69 &--0.745(0.225) &--0.15(0.04) \\
     & 550 &  --0.73 & --3.38 & 1.69 &--1.030(0.200) &--0.57(0.02) \\
\hline
\hline
\end{tabular}
\end{table}

\subsection{Single-nucleon momentum distributions and corresponding integrated SRC probabilities} 
\label{subsec:single-n} 

The 1N momentum distributions for a particular nucleon ($p$ or $n$)
with momentum ${\bf k}$ in $^3$He are defined as                         
\begin{equation}
  n^{p/n}(k) = \frac{1}{2} \int d{\hat{\bf k}}\; d{\bf q}\,
  \Psi^\dagger({\bf k},{\bf q}) P_{p/n}\Psi({\bf k},{\bf q})\ , 
  \label{eq:1b-pn}
\end{equation}
where we have fixed the permutation to be $p=1$, i.e.\ the particular
nucleon is fixed to be particle $1$, and therefore ${\bf k}={\bf k}_{p=1}$
and ${\bf q}={\bf q}_{p=1}$, in the notation of
Eq.~(\ref{eq:jacobip}). Furthermore, $P_{p/n}$
is the proton/neutron projection operator acting on particle 1.
With this definition, the 1N momentum distributions
are normalized as
\begin{equation}
  4\pi\int k^2\, dk\, n^{n/p}(k) =1
  \label{eq:norm-1b} \ .
\end{equation}
We have verified that Eqs.~(\ref{eq:1b-pn}) and~(\ref{eq:norm-1b})
are consistent with those of Ref.~\cite{Alv13}.

\begin{figure}[tbh!] 
\centering         
\scalebox{0.5}{\includegraphics{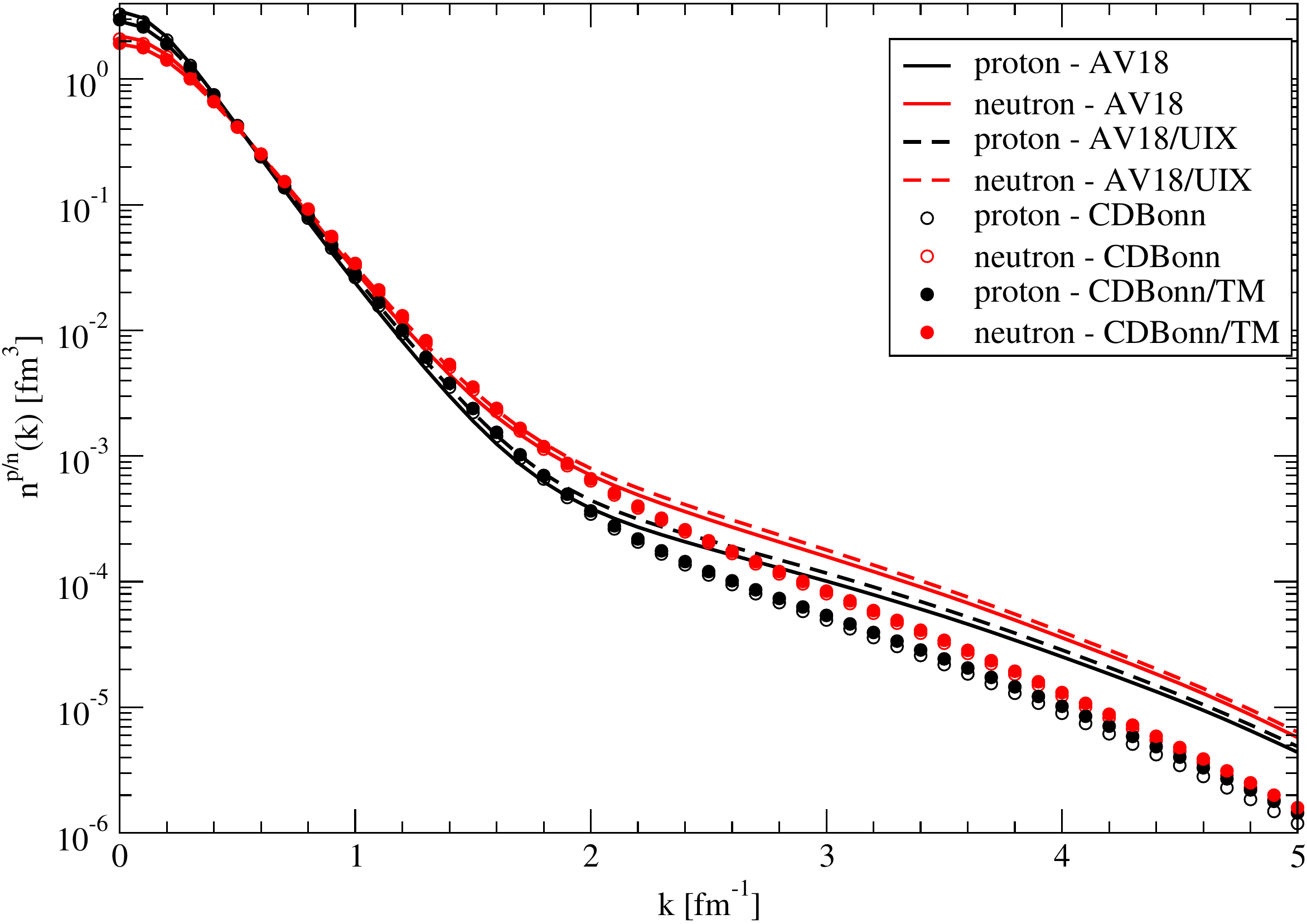}}
\caption{(Color online) The 1N momentum
      distributions  $n^{p/n}(k)$, calculated using the
      AV18 and CDBonn 2N interaction, AV18/UIX and CDBonn/TM
      2N and 3N interaction models.}
\label{fig:1bmd-2nf-phen}
\end{figure}

We have first calculated the 1N momentum distributions using the
AV18~\cite{av18} or CDBonn~\cite{CD} phenomenological potentials,
with and without the 3N force (UIX~\cite{UIX}                      
or TM~\cite{TM} for AV18 or CDBonn, respectively).                   
The results are shown in Fig.~\ref{fig:1bmd-2nf-phen}. From those, 
we conclude that 3N force contributions are small, and only noticeable 
for $k\ge 2$ fm$^{-1}$. On the contrary, potential-model dependence         
is large in the range $k\ge 2$ fm$^{-1}$, an aspect which will be a recurrent theme
throughout this paper. To avoid an excessively cumbersome presentation, we are not showing the results of
Ref.~\cite{Alv13} and Ref.~\cite{Wir_web}, obtained using
AV18 HH and AV18/UIX Variational Monte Carlo (VMC)
wave functions. However, 
we have verified that we are in agreement with 
Refs.~\cite{Alv13,Wir_web}, with small differences only
in the high-$k$ tail of the distributions. Comparison between
our results and those of Refs.~\cite{Alv13,Alv16,Wir_web}
will be shown in the case of the back-to-back 2N momentum
distribution (see below).

\begin{figure}[tbh!] 
\centering         
\scalebox{0.35}{\includegraphics{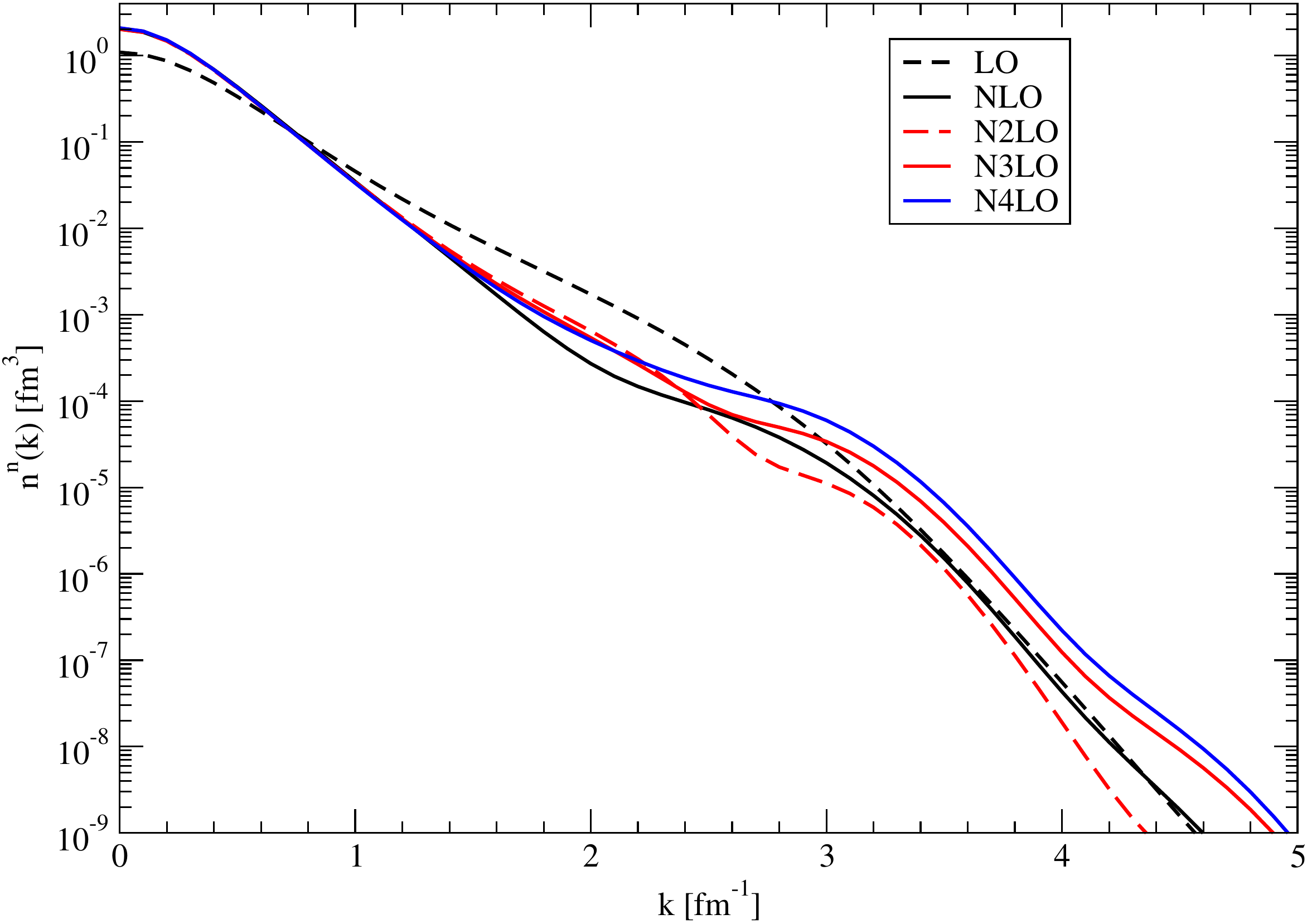}}
\scalebox{0.35}{\includegraphics{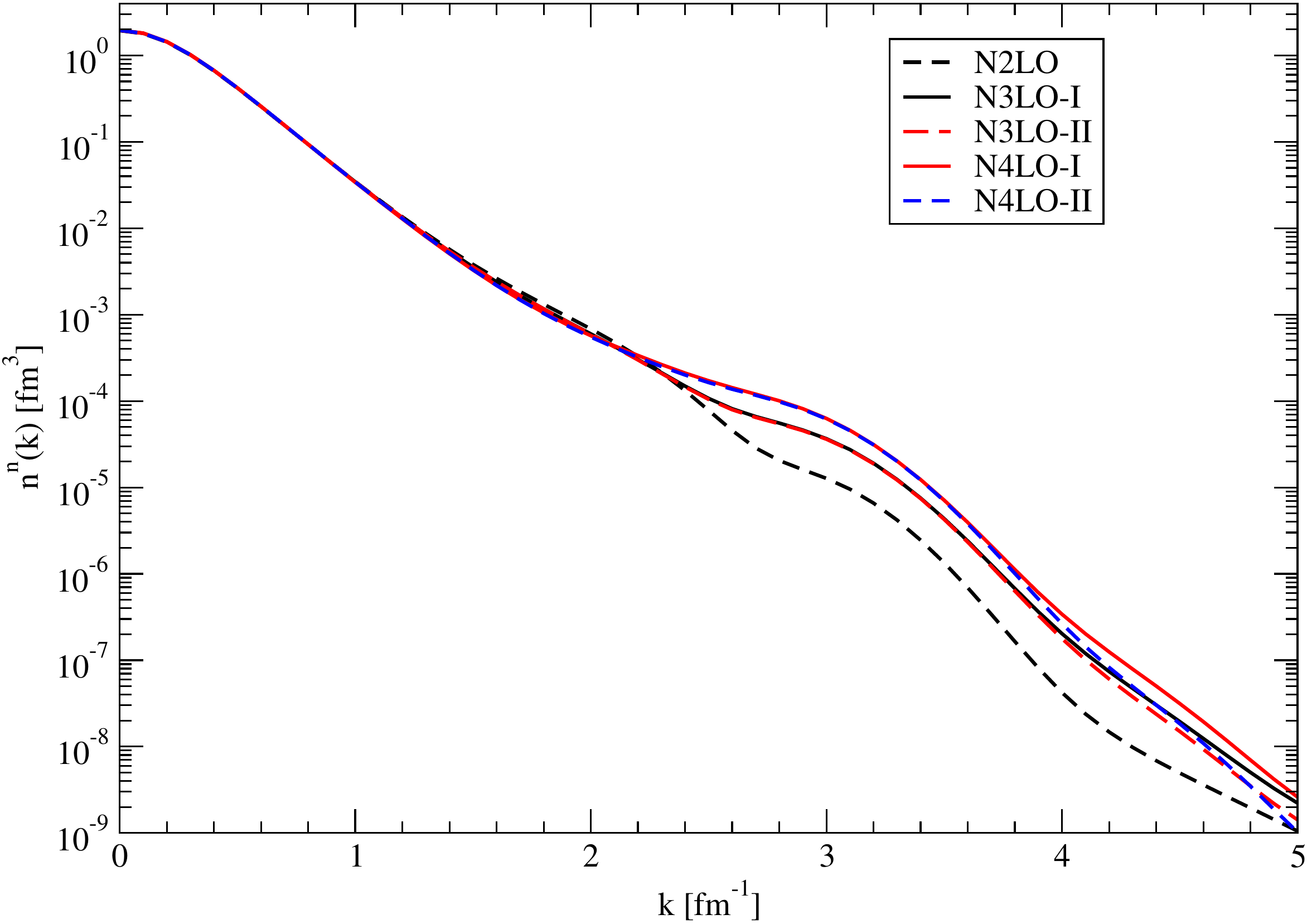}}
\caption{(Color online) The neutron momentum
  distributions  $n^{n}(k)$, calculated using
  only 2N (left panel) and 2N+3N (right panel) chiral interactions,
  with $\Lambda=500$ MeV.
  The different chiral order are labelled as in the text. In particular,
  on the right panel, we have indicated with ``I'' and ``II''
  the results
  obtained using the LECs of Table~\ref{tab:cdceI}
  and~\ref{tab:cdceII}, respectively.}
\label{fig:1bnmd-chiral}
\end{figure}

\begin{figure}[tbh!] 
\centering         
\scalebox{0.35}{\includegraphics{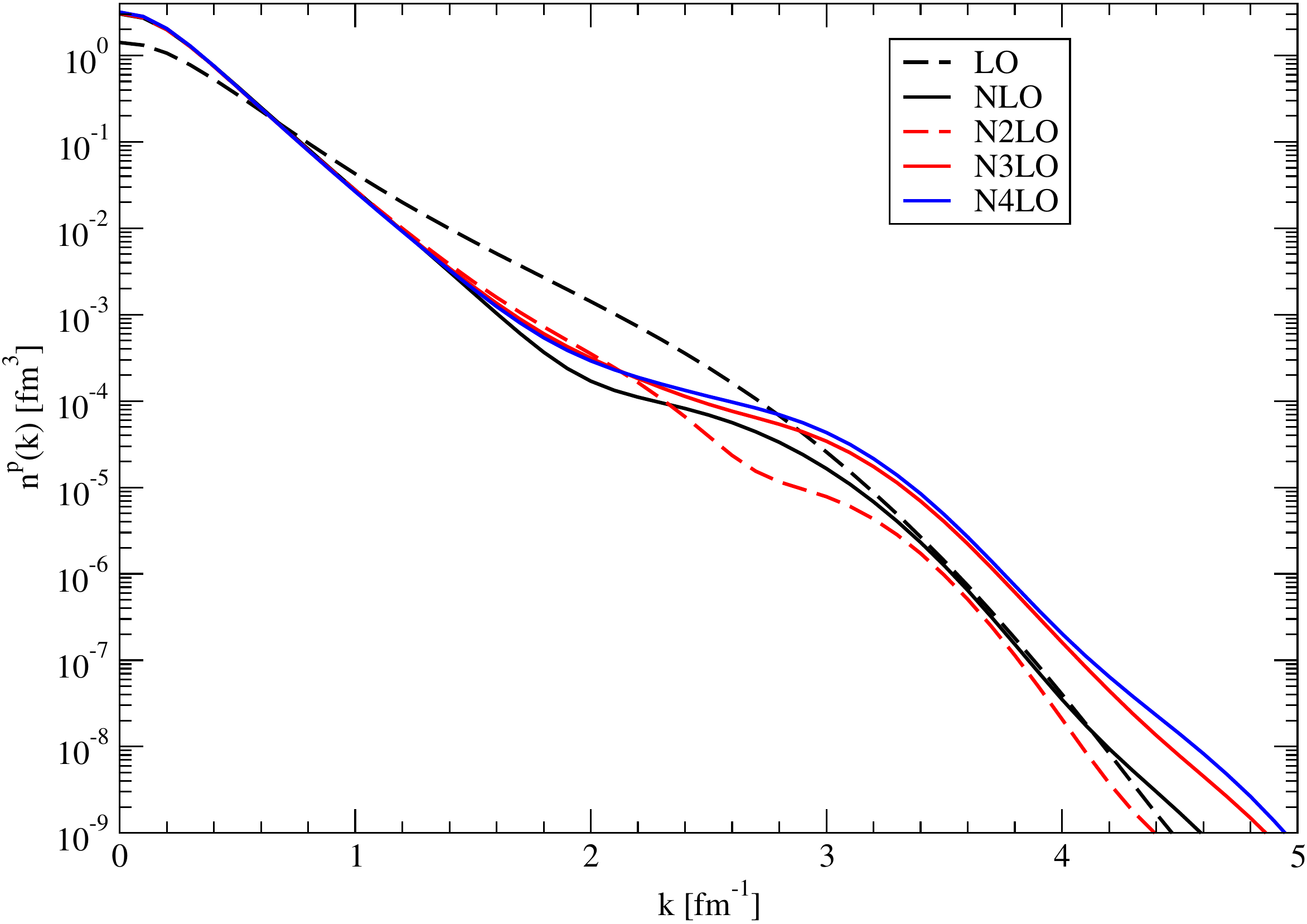}}
\scalebox{0.35}{\includegraphics{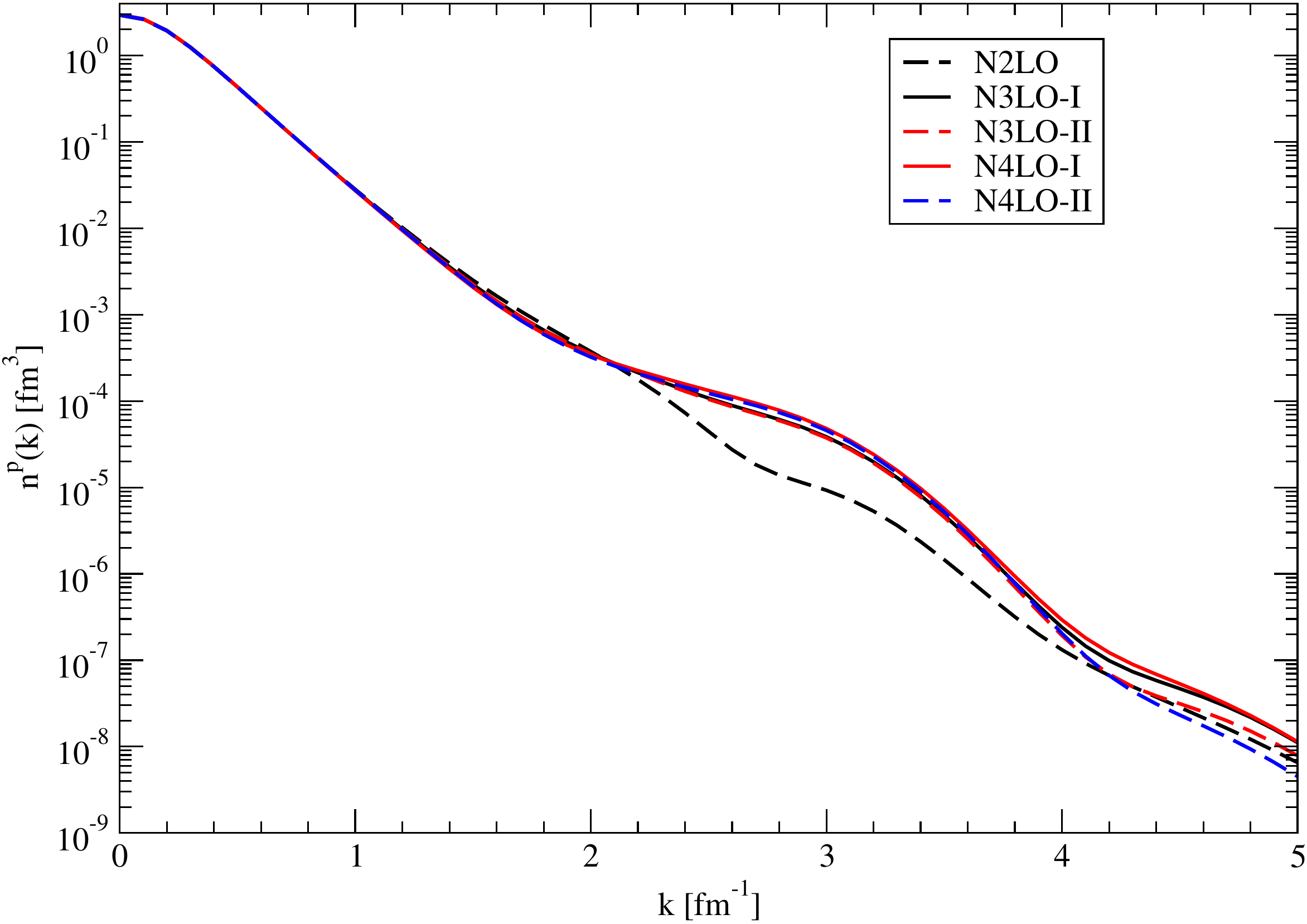}}
\caption{Same as Fig.~\ref{fig:1bnmd-chiral} but for the
  proton momentum distributions $n^{p}(k)$.}
\label{fig:1bpmd-chiral}
\end{figure}

We then move to study the 1N momentum distributions
using chiral potentials~\cite{EMN}.
In Figs.~\ref{fig:1bnmd-chiral} and~\ref{fig:1bpmd-chiral}
we present the neutron and proton 1N momentum
distribution obtained with only 2N forces at LO,
NLO, N2LO, N3LO and N4LO on the left panel, and
adding the 3N force, with LECs 
from Table~\ref{tab:cdceI} (model I) or~\ref{tab:cdceII}
(model II). These 1N momentum distributions
are calculated with cutoff value fixed at $\Lambda=500$ MeV.
The figure shows that, for small values of $k$, all predictions at NLO and
higher orders are quite similar. Overall, differences 
between the N3LO and N4LO curves are small enough              
to suggest a reasonable convergence pattern.  
The 3N force
contributions are found again to be very small, and therefore       
the differences between the predictions from 
model I and model II for the 3N force are even smaller.

\begin{figure}[tbh!] 
\centering         
\scalebox{0.7}{\includegraphics{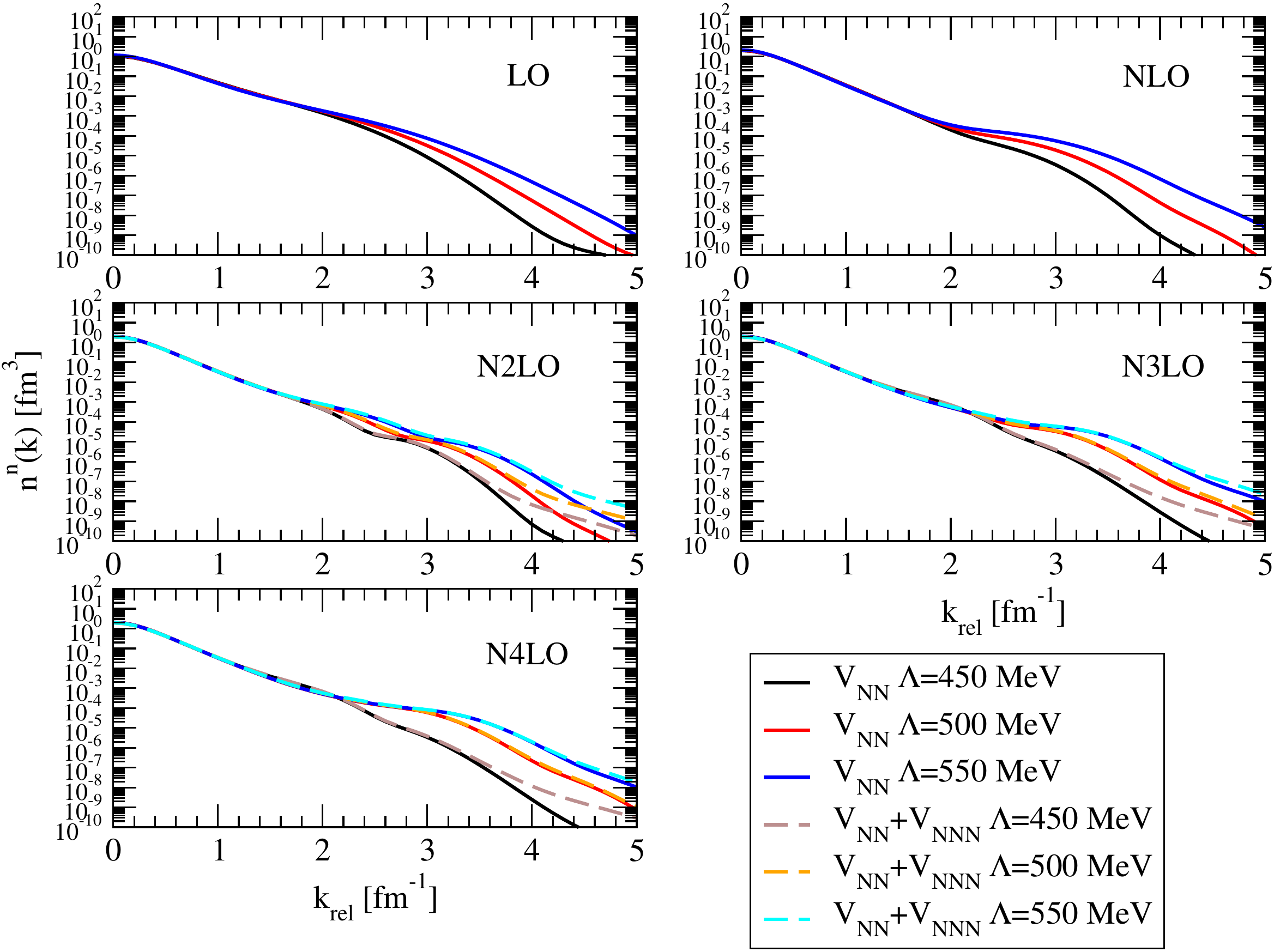}}
\caption{(Color online) The neutron momentum distributions
  $n^{n}(k)$, calculated using
  only 2N (solid lines) and 2N+3N (dashed lines) chiral interactions,
  at different chiral order and for three values of the  
  cutoff $\Lambda= 450, 500, 550$ MeV. The LECs of the
  3N interaction are those of Table~\ref{tab:cdceII}.}
\label{fig:1bnmd-allL}
\end{figure}

\begin{figure}[tbh!] 
\centering         
\scalebox{0.7}{\includegraphics{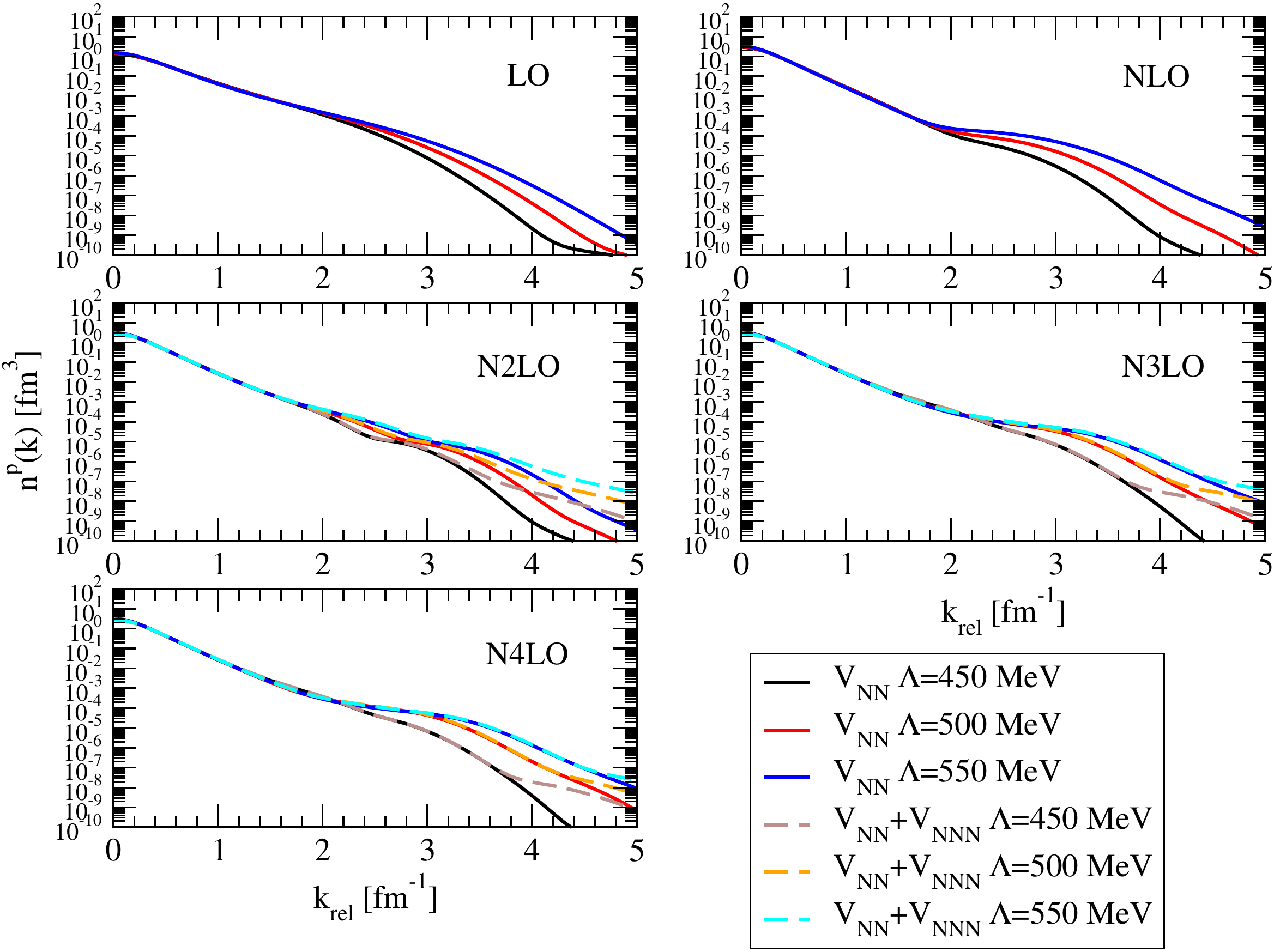}}
\caption{Same as Fig.~\ref{fig:1bnmd-allL} but for the
  proton momentum distributions $n^{p}(k)$.}
\label{fig:1bpmd-allL}
\end{figure}

The 1N momentum distributions $n^n(k)$ and $n^p(k)$ calculated with
and without 3N interaction, at different chiral orders and
for different values of the cutoff $\Lambda$, are shown in
Figs.~\ref{fig:1bnmd-allL} and~\ref{fig:1bpmd-allL}, respectively.
By inspection of the figures, we can see that
cutoff dependence appears comparable at all orders. Naturally, sensitivity is 
more pronounced in the high $k_{rel}$ region, where larger values of the cutoff 
produce ``harder" distributions. Note, also, that cutoff-induced differences
become
noticeable where the distributions are reduced by about 5 or 6 orders of
magnitude from their
maximum values. 

Finally we consider the integrated probabilities,
defined as
\begin{equation}
  {\cal{P}}^{p/n}(k^-)=4\pi\int_{k^-}^\infty n^{p/n}(k)\,k^2\,dk
  \label{eq:Ppn}
\end{equation}
as in Table II of Ref.~\cite{Alv13}. The results obtained
with the 2N and 2N+3N phenomenological potentials are
listed in Table~\ref{tab:1bint-phen}. Those obtained using
the chiral potentials are presented in Table~\ref{tab:1bint-chiral}.
In both tables, we have first calculated
${\cal{P}}^{p/n}(k^-=0)$, in order to verify that the 1N momentum distributions
are properly normalized. Note that in our integration, the
upper limit of the integral is in fact 5 fm$^{-1}$. Therefore,
the difference of ${\cal{P}}^{p/n}(k^-=0)$ from unity
(see Eq.~(\ref{eq:norm-1b}))
gives an indication of the importance of the tail of the momentum
distribution. By inspection of the tables we can see that
${\cal{P}}^{p/n}(k^-=0)\simeq 1$ within 0.2--0.3\%.
A comparison of the results of Table~\ref{tab:1bint-phen}
and~\ref{tab:1bint-chiral} shows again a remarkable model dependence.
The results of Table~\ref{tab:1bint-chiral} show also a satisfactory
order-by-order convergence.
\begin{table}[t]
  \caption{\label{tab:1bint-phen} The integrated $p/n$ 
    probabilities, as defined in Eq.~(\ref{eq:Ppn}),
    obtained with the various phenomenological
    potential models, i.e.\ AV18, CDBonn,
    AV18/UIX and CDBonn/TM, varying the integration lower bound
    $k^-$, given in fm$^{-1}$. We report also the values of
    Table II of Ref.~\cite{Alv13}, obtained with the AV18 potential
    model.}
\begin{tabular}{c|cc|cc}
    \hline
    \hline
  & \multicolumn{2}{c|}{$P^n(k^-)$} & \multicolumn{2}{|c}{$P^p(k^-)$}
    \\ 
\hline
 & $k^-=0.0$ & $k^-=1.5$ & $k^-=0.0$ & $k^-=1.5$\\
\hline
AV18     & 0.997 & 0.068 & 0.997 & 0.041 \\
AV18/UIX & 0.997 & 0.077 & 0.998 & 0.048 \\
\hline
CDBonn    & 0.998 & 0.052 & 0.998 & 0.031 \\
CDBonn/TM & 0.998 & 0.054 & 0.999 & 0.033 \\
\hline
Ref.~\cite{Alv13} & 0.999 & 0.067 & 1.000 & 0.041 \\
\hline
\hline
  \end{tabular}

\end{table}
\begin{table}[t]
  \caption{\label{tab:1bint-chiral} Same as Table~\ref{tab:1bint-phen}
    but
    obtained with chiral
    potentials, at different chiral orders, for different
    values of the cutoff $\Lambda$, 450, 500 and 550 MeV,
    also with the inclusion of the 3N force
    (lines labelled N2LO/N2LO at N2LO, N3LO/N3LO-I and
    N3LO/N3LO-II at N3LO, N4LO/N4LO-I and N4LO/N4LO-II at N4LO).
    The labels ``I'' and ``II'' refer to the LECs of Table~\ref{tab:cdceI}
    and~\ref{tab:cdceII}, respectively. }
\begin{tabular}{c|ccc|ccc|ccc|ccc}
    \hline
    \hline
  & \multicolumn{3}{c|}{$P^n(k^-=0.0)$} & \multicolumn{3}{c|}{$P^p(k^-=0.0)$}
  & \multicolumn{3}{c|}{$P^n(k^-=1.5)$} & \multicolumn{3}{c}{$P^p(k^-=1.5)$}
    \\ 
\hline
Model/$\Lambda$ [MeV] & 450 & 500 & 550 & 450 & 500 & 550
& 450 & 500 & 550 & 450 & 500 & 550 \\
\hline
LO
& 1.000 & 1.000 & 0.999
& 1.000 & 0.999 & 0.999
& 0.090 & 0.105 & 0.113
& 0.076 & 0.089 & 0.095\\
NLO
& 0.999 & 0.999 & 0.999 
& 0.999 & 0.999 & 0.998
& 0.020 & 0.025 & 0.033
& 0.013 & 0.016 & 0.023\\ 
\hline
N2LO
& 0.999 & 0.999 & 0.999
& 0.999 & 0.999 & 0.999
& 0.033 & 0.040 & 0.046
& 0.020 & 0.024 & 0.027 \\
N2LO/N2LO
& 0.999 & 0.999 & 0.999
& 0.999 & 0.999 & 0.999
& 0.033 & 0.040 & 0.046
& 0.020 & 0.024 & 0.027 \\
\hline
N3LO
& 0.999 & 0.999 & 0.999
& 0.999 & 0.999 & 0.999
& 0.042 & 0.038 & 0.041
& 0.025 & 0.025 & 0.026 \\
N3LO/N3LO-I  
& 0.999 & 0.999 & 0.999
& 0.999 & 0.999 & 0.999
& 0.045 & 0.041 & 0.045
& 0.027 & 0.028 & 0.030 \\
N3LO/N3LO-II
& 0.999 & 0.999 & 0.999
& 0.999 & 0.999 & 0.999
& 0.045 & 0.041 & 0.045
& 0.027 & 0.027 & 0.029 \\
\hline
N4LO
& 0.999 & 0.999 & 0.999
& 0.999 & 0.999 & 0.999
& 0.041 & 0.039 & 0.043
& 0.024 & 0.025 & 0.026 \\
N4LO/N4LO-I  
& 0.999 & 0.999 & 0.999
& 0.999 & 0.999 & 0.999
& 0.043 & 0.043 & 0.048
& 0.026 & 0.028 & 0.030\\
N4LO/N4LO-II 
& 0.999 & 0.999 & 0.999 
& 0.999 & 0.999 & 0.999
& 0.043 & 0.042 & 0.046
& 0.026 & 0.027 & 0.028 \\
\hline
\hline
\end{tabular}
\end{table}

A glance at Figs.~\ref{fig:1bmd-2nf-phen}
--~\ref{fig:1bpmd-allL},
reveals characteristic differences between the qualitative features of the 
chiral predictions as compared to the phenomenological and
meson-theoretic ones. This is due to the polynomial structure of the 
(short-range) contact terms used in the construction of the chiral potentials,
combined with the exponential regulator function 
\begin{equation}
f(p',p) = \exp[-(p'/\Lambda)^{2n} - (p/\Lambda)^{2n}] \; .
\label{reg}
\end{equation}
In the meson-theoretic potentials, the short range is described by heavy-meson
exchanges represented by 
Yukawa functions of heavy-meson masses. On the other hand, the
phenomenological AV18 potentials uses 
a Woods-Saxon function to provide the short-range core.
(Heavy mesons, of course, have no place in 
chiral EFT.) 
 Overall, the chiral predictions fall off at a faster rate as compared to the 
phenomenological ones. This is to be expected from the ``softer" nature of the chiral potentials.

\subsection{Two-nucleon momentum distributions and corresponding integrated SRC probabilities} 
\label{subsec:2N} 

The 2N momentum distribution of the $N_1N_2$ pair,
with $N_1N_2=np$ or $pp$, as
a function of their relative momentum $k_{rel}$, is defined as
\begin{equation}
  n^{N_1N_2}(k_{rel},K_{c.m.}) =
  \int d{\hat{\bf k}}_{rel}
  \int d{\hat{\bf k}}_{c.m.}
  \Psi^{\dagger}({\bf k}_{rel},{\bf K}_{c.m.})
  P_{N_1N_2}\Psi({\bf k}_{rel},{\bf K}_{c.m.}) \ ,
\label{eq:2NMD-1}
\end{equation}
where $P_{N_1N_2}$ is the projection operator on the $N_1N_2$ pair.
Note that we have introduced the definitions 
\begin{eqnarray}
  {\bf k}_{rel}&=&-\frac{\sqrt{2}}{2} {\bf q}_{p=1} \ , \nonumber \\
  {\bf K}_{c.m.}&=&-\sqrt{\frac{2}{3}} {\bf k}_{p=1} \ , \label{eq:2b-corr}
\end{eqnarray}
that is, we have chosen the pair $N_1N_2$ to contain particles $2,3$. 
In the following, we will focus on the so-called back-to-back (BB)
2N momentum distributions,
i.e.\  $n^{N_1N_2}(k_{rel},K_{c.m.}=0)$, and on the
$K_{c.m.}$-integrated 2N momentum distributions,
i.e.\
\begin{equation}
  n^{N_1N_2}(k_{rel}) = 4\pi
  \int _{0}^{K^+_{c.m.}} K^2_{cm}\,dK_{cm}   n^{N_1N_2}(k_{rel},K_{c.m.}) \ .
\label{eq:2NMD-2}
\end{equation}
The upper limit of the $K_{c.m.}$-integration restricts the values of
$K_{c.m.}$ to a limited range, approximately $K^+_{c.m.}\simeq 1.0-1.5$ fm$^{-1}$.
This is because, in the SRC model (as opposed to the mean-field model),
one considers highly 
correlated $N_1N_2$ pairs with small center-of-mass
momentum~\cite{Alv13,Alv16}.
The integrations of Eqs.~(\ref{eq:2NMD-1}) and~(\ref{eq:2NMD-2})
have been performed numerically with the Van der Corput sequence~\cite{VDC}
and we have verified that our results are stable with the increasing
value of Van der Corput points of integrations. Typically, 50 000
points are enough for converged results.

\begin{figure}[tbh!] 
\centering         
\scalebox{0.35}{\includegraphics{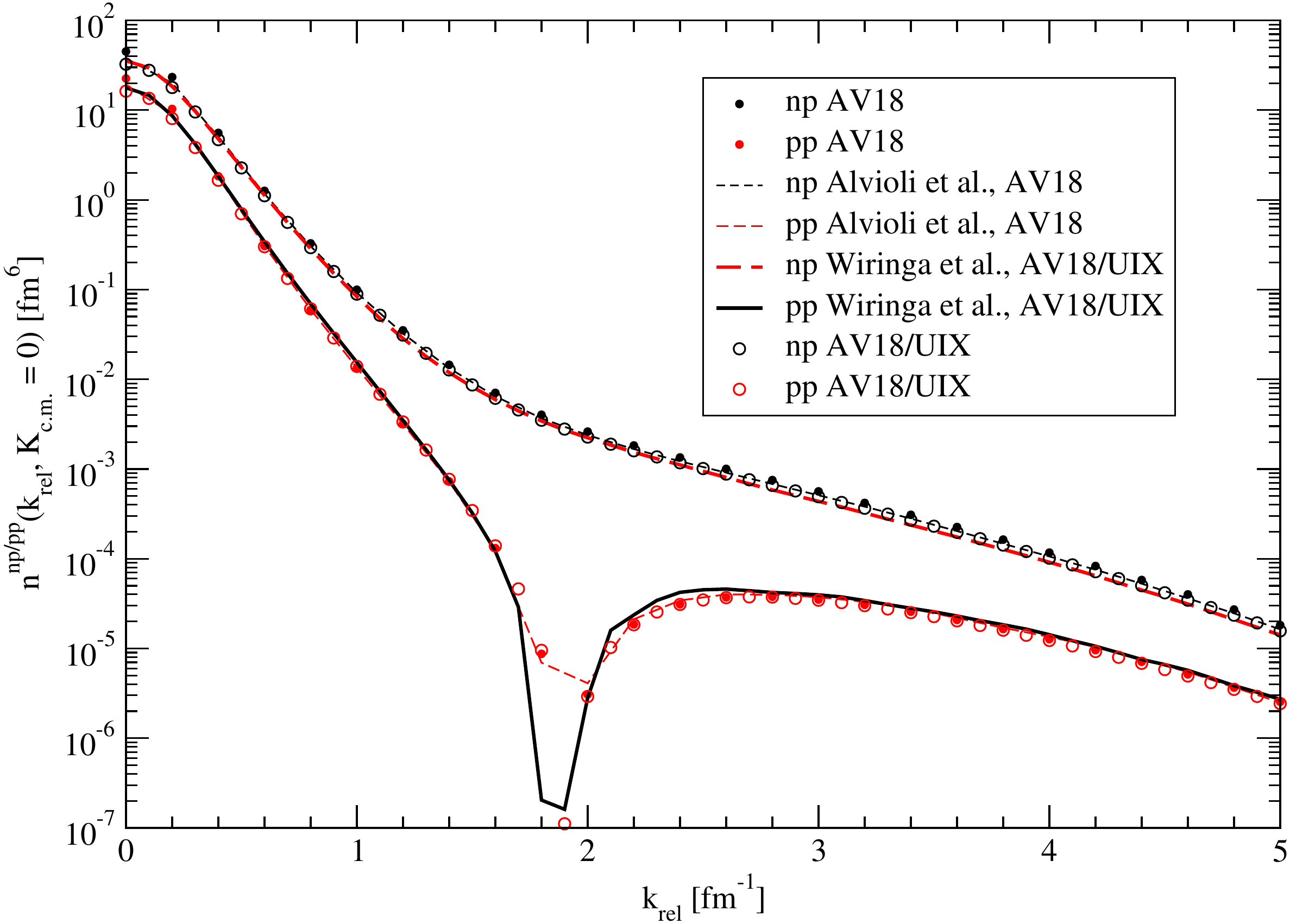}}
\caption{(Color online) The 2N momentum
      distributions  $n^{np/pp}(k_{rel},K_{c.m.}=0)$, calculated using the
      AV18 and the AV18/UIX interaction models. The
      present results are shown as dots, while the previous
      studies of Refs.~\cite{Alv13,Alv16} and Ref.~\cite{Wir_web} are
shown as dashed or continuous lines.}
\label{fig:2bmd_av18uix-kcm0}
\end{figure}

First we calculate the 2N momentum distributions using 
the AV18~\cite{av18} phenomenological potential,
with and without the UIX phenomenological 3N force~\cite{UIX},
in order to compare with results available in the
literature~\cite{Alv13,Alv16,Wir_web}.
The comparison presented in                                             
Fig.~\ref{fig:2bmd_av18uix-kcm0} shows that 
we are able to reproduce
the results of previous investigations for                            
$n^{N_1N_2}(k_{rel},K_{c.m.}=0)$,
but we have verified a similar degree of agreement   
also for $n^{N_1N_2}(k_{rel})$.
Furthermore, we see that the 3N force
contribution is quite small, an observation which will 
be confirmed throughout the paper.                         

\begin{figure}[tbh!] 
\centering         
\scalebox{0.3}{\includegraphics{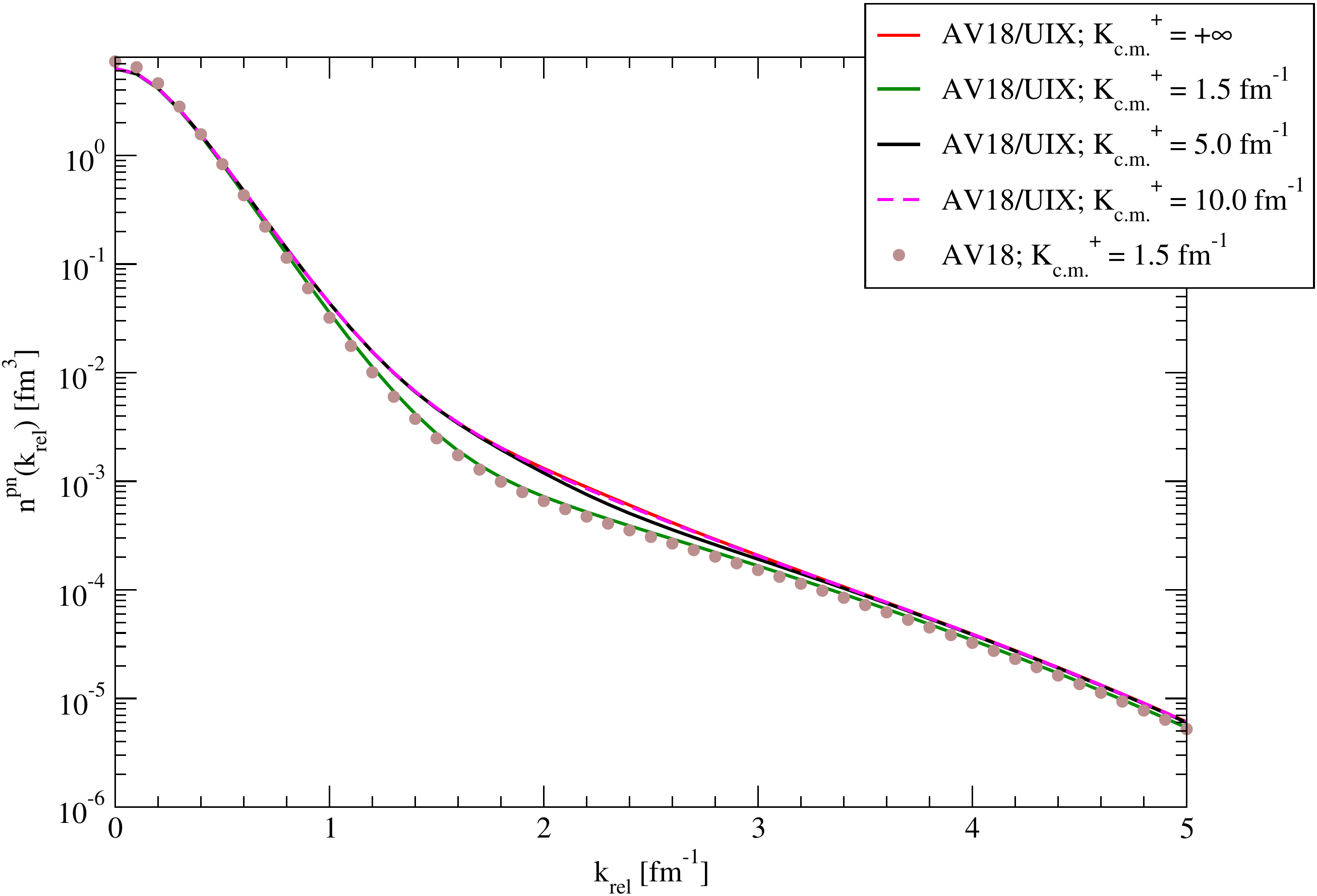}}
\scalebox{0.3}{\includegraphics{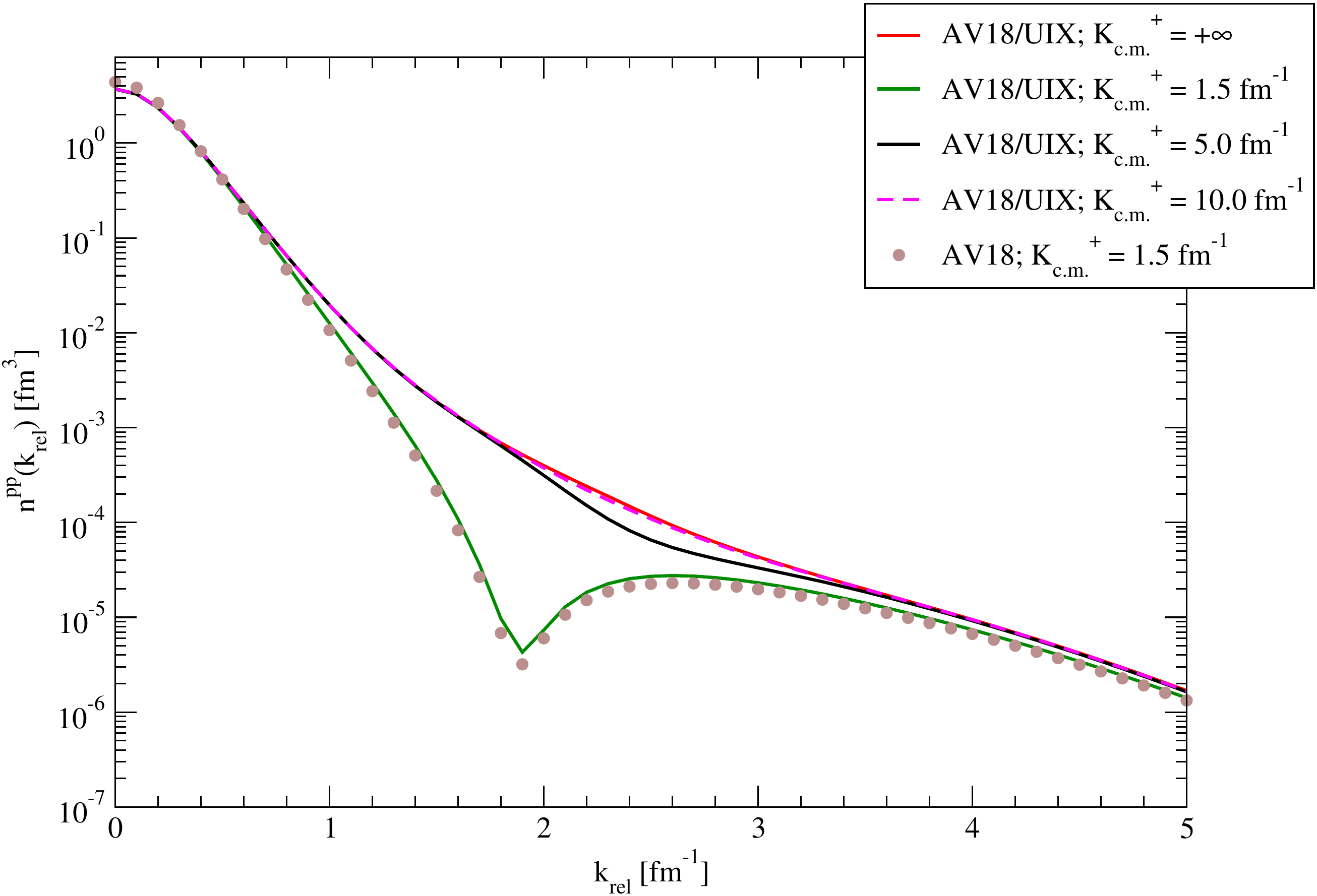}}
\caption{(Color online) The 2N momentum distributions
  $n^{np}(k_{rel})$ (left panel) and
  $n^{pp}(k_{rel})$ (right panel) as function
  of $K_{c.m.}^+$ (see Eq.~(\ref{eq:2NMD-2})),
  calculated using the AV18/UIX potential. For $K_{c.m.}^+=1.5$ fm$^{-1}$
  we show also the results obtained with the AV18 2N only potential.} 
\label{fig:2bmd_av18uix-kcm+}
\end{figure}

We then move to the $n^{np/pp}(k_{rel})$ 2N momentum distribution 
as function of $K_{c.m.}^+$ (see Eq.~(\ref{eq:2NMD-2})). The results for the
AV18/UIX are shown in Fig.~\ref{fig:2bmd_av18uix-kcm+}, from which we can conclude
that contributions from $K_{c.m.}^+$ 
larger than approximately 5 fm$^{-1}$ are not significant. 
We also note, in passing, that for $K_{c.m.}^+=1.5$ fm$^{-1}$, the AV18 and AV18/UIX results
are very close to each other. 

\begin{figure}[tbh!] 
\centering         
\scalebox{0.35}{\includegraphics{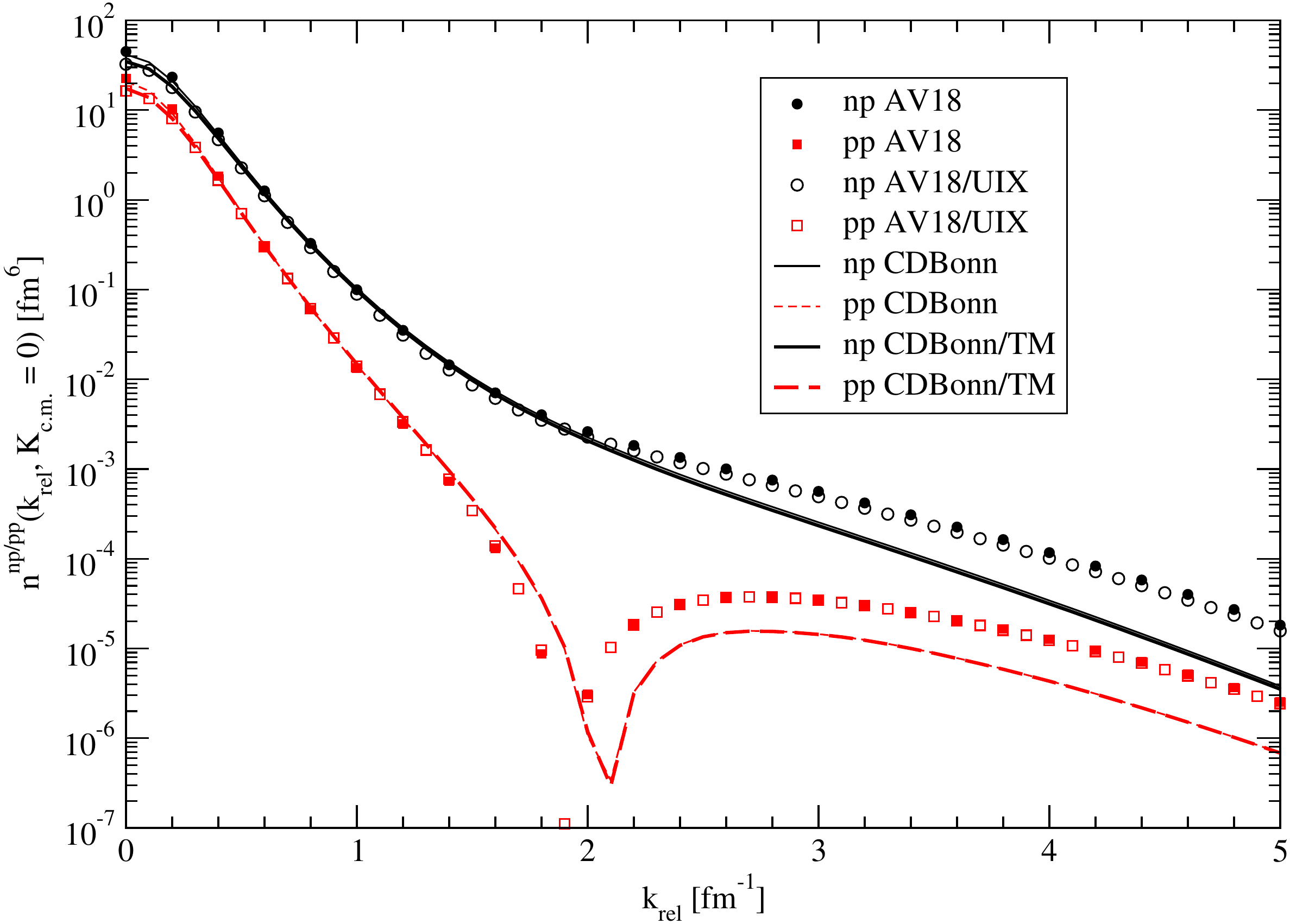}}
\caption{(Color online) The 2N momentum
      distributions  $n^{np/pp}(k_{rel},K_{c.m.}=0)$, calculated using the
      AV18, AV18/UIX, CDBonn and CDBonn/TM 2N and 3N interaction models.
      The thin and think lines are on top of each other and can barely be
      distinguished.} 
\label{fig:2bmd_cdbtm-kcm0}
\end{figure}

\begin{figure}[tbh!] 
\centering         
\scalebox{0.3}{\includegraphics{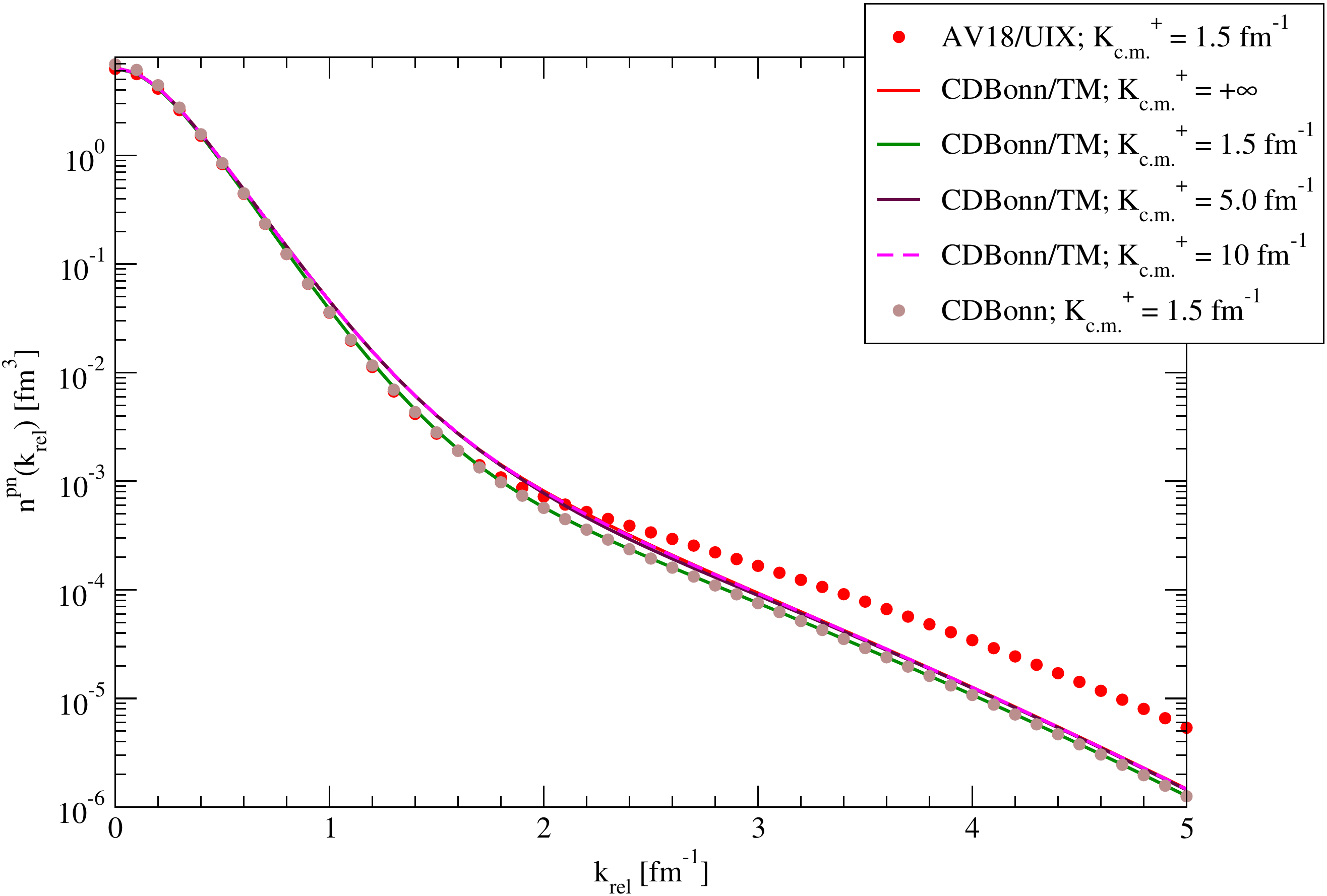}}
\scalebox{0.3}{\includegraphics{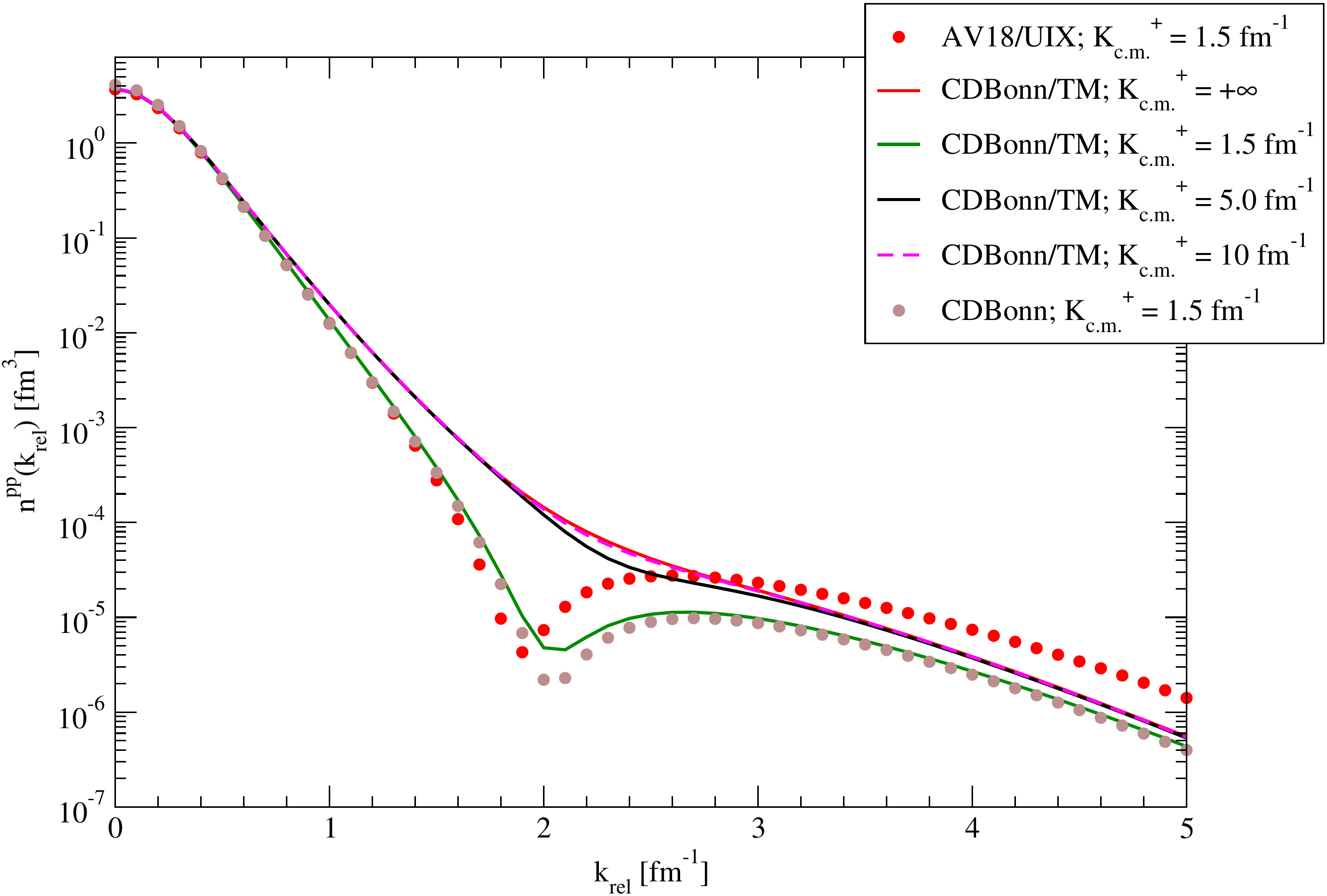}}
\caption{(Color online) The 2N momentum distributions
  $n^{np}(k_{rel})$ (left panel) and
  $n^{pp}(k_{rel})$ (right panel) as function
  of $K_{c.m.}^+$ (see Eq.~(\ref{eq:2NMD-2})),
  calculated using the CDBonn/TM potential. For $K_{c.m.}^+=1.5$ fm$^{-1}$
  we show also the results obtained with the CDBonn 2N only potential
  and the AV18/UIX potential, already presented in
  Fig.~\ref{fig:2bmd_av18uix-kcm+}.} 
\label{fig:2bmd_cdbtm-kcm+}
\end{figure}

Next we explore the model-dependence of the 2N momentum distributions,
by repeating the calculations using the CDBonn potential without
or with the TM 3N force. In Figs.~\ref{fig:2bmd_cdbtm-kcm0}
and~\ref{fig:2bmd_cdbtm-kcm+} we show the results for the
$n^{np/pp}(k_{rel},K_{c.m.}=0)$ and for the $n^{np/pp}(k_{rel})$
as function of $K_{c.m.}^+$, respectively.
The figures reveal that:
(i) the results with CDBonn/TM and those with AV18/UIX are substantially different from each other, 
 especially in the high-$k_{rel}$ tails, confirming                             
what we mentioned earlier while recalling the findings 
of Ref.~\cite{src2015};                                  
(ii) the 3N force contributions are again barely appreciable
on the plot (which are on a logarithmic scale); (iii) the $K_{c.m.}^+$-dependence
in the CDBonn/TM case is very similar to the one seen in the AV18/UIX case.

An important issue in the considerations of SRC is the behavior of
$n^{np}(k_{rel})$ in nuclei as compared with       
the same quantity in the deuteron
($n^{np}_d(k_{rel})$). This is because a highly correlated
$np$ pair 
in a nucleus is expected to exhibit a behavior similar to the pair in
the deuteron.
We proceed to calculate the integrated
SRC-probabilities defined as
\begin{eqnarray}
  N_{N_1N_2}^{BB}&=&
  4\pi\int_0^\infty n^{N_1N_2}(k_{rel},K_{c.m.}=0) k_{rel}^2\,
  dk_{rel} \ , \label{eq:n-bb}\\
  N_{N_1N_2}^{SRC,BB}&=&
  4\pi\int_{k_{rel}^-}^\infty n^{N_1N_2}(k_{rel},K_{c.m.}=0) k_{rel}^2\,
  dk_{rel} \ , \label{eq:n-src-bb} \\
  N_{N_1N_2}^{SRC}(k_{rel}^-)&=&
  4\pi\int_{k_{rel}^-}^\infty n^{N_1N_2}(k_{rel}) k_{rel}^2\,
  dk_{rel} \ , \label{eq:n-src} \\
  N_{N_1N_2}&=&
  4\pi\int_0^\infty n^{N_1N_2}(k_{rel}) k_{rel}^2\,
  dk_{rel} \equiv N_{N_1N_2}^{SRC}(k_{rel}^-=0) \ , \label{eq:n}
\end{eqnarray}
where we have used $k_{rel}^-=1.5$ fm$^{-1}$. These equations are
the same as in Ref.~\cite{Alv13}. 
For convenience, we will continue to refer to these integrated quantities as 
probabilities. A more accurate description of, for instance, $N_{N_1N_2}^{BB}$ would be 
the number of back-to-back $N_1N_2$ pairs after integration of the pair relative momentum. 

The results for the different potential models used so far are shown in
Table~\ref{tab:nnn_phen}, from which we can conclude that
the 3N force
contributions are small also for the integrated quantities. However,      
model-dependence is strong, especially for $N^{SRC,BB}$
and $N^{SRC}(k_{rel}^-)$. This large model-dependence might have      
impact on the extraction of SRC probabilities from $(e,e^\prime p)$
experiments, if not properly taken into account.
\begin{table}[t]
  \caption{\label{tab:nnn_phen} The integrated $np$ and $pp$
    SRC-probabilities, as defined in Eqs.~(\ref{eq:n-bb})--(\ref{eq:n}),
    obtained with the various potential models, i.e.\ AV18, CDBonn,
    AV18/UIX and CDBonn/TM. For the $np$ case, we report in the last two lines
    labelled AV18 - d and CDBonn - d
    the deuteron results of Ref.~\cite{src2015}.}
  \begin{tabular}{c|cccc|cccc}
    \hline
    \hline
    &\multicolumn{4}{|c|}{$N_1N_2=np$} & \multicolumn{4}{|c}{$N_1N_2=pp$} \\ 
\hline
& $N^{BB}$ & $N^{SRC,BB}$ & $N^{SRC}(k_{rel}^-)$ & $N$ 
& $N^{BB}$ & $N^{SRC,BB}$ & $N^{SRC}(k_{rel}^-)$ & $N$ \\
\hline
AV18      & 6.922   & 0.241      & 0.093    & 1.997
          & 2.194   & 0.009      & 0.026    & 0.998 \\
AV18/UIX  & 5.751   & 0.210      & 0.106    & 1.997
          & 1.897   & 0.009      & 0.031    & 0.999 \\
\hline 
CDBonn    & 6.552   & 0.171      & 0.060    & 1.997
          & 2.078   & 0.005      & 0.012    & 0.999 \\
CDBonn/TM & 5.931   & 0.157      & 0.063    & 1.998
          & 1.924   & 0.005      & 0.014    & 0.998 \\
\hline
AV18 - d  &         &            & 0.042      & \\
CDBonn - d&         &            & 0.032      & \\
\hline
\hline
  \end{tabular}
\end{table}

\begin{figure}[tbh!] 
\centering         
\scalebox{0.35}{\includegraphics{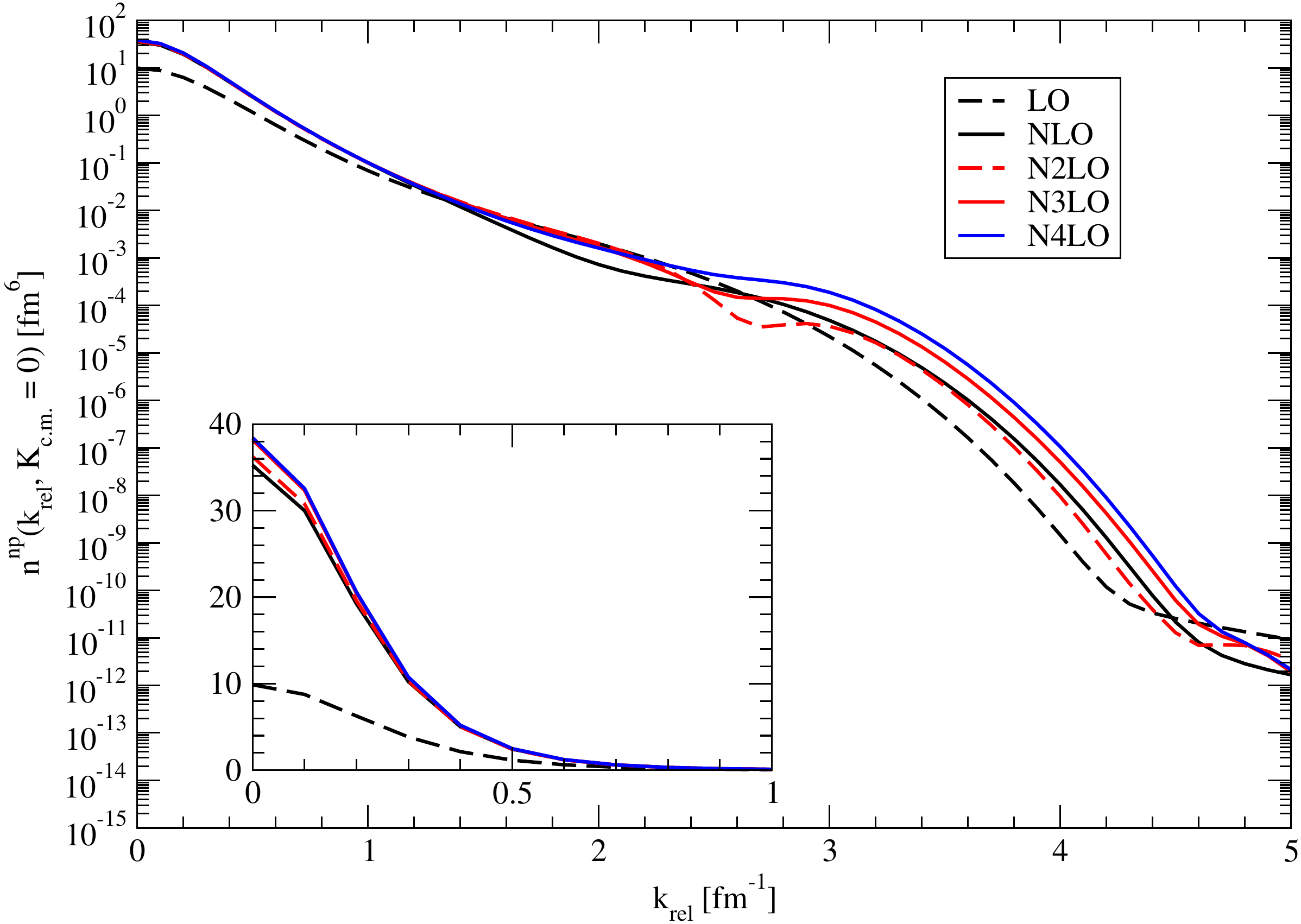}}
\scalebox{0.35}{\includegraphics{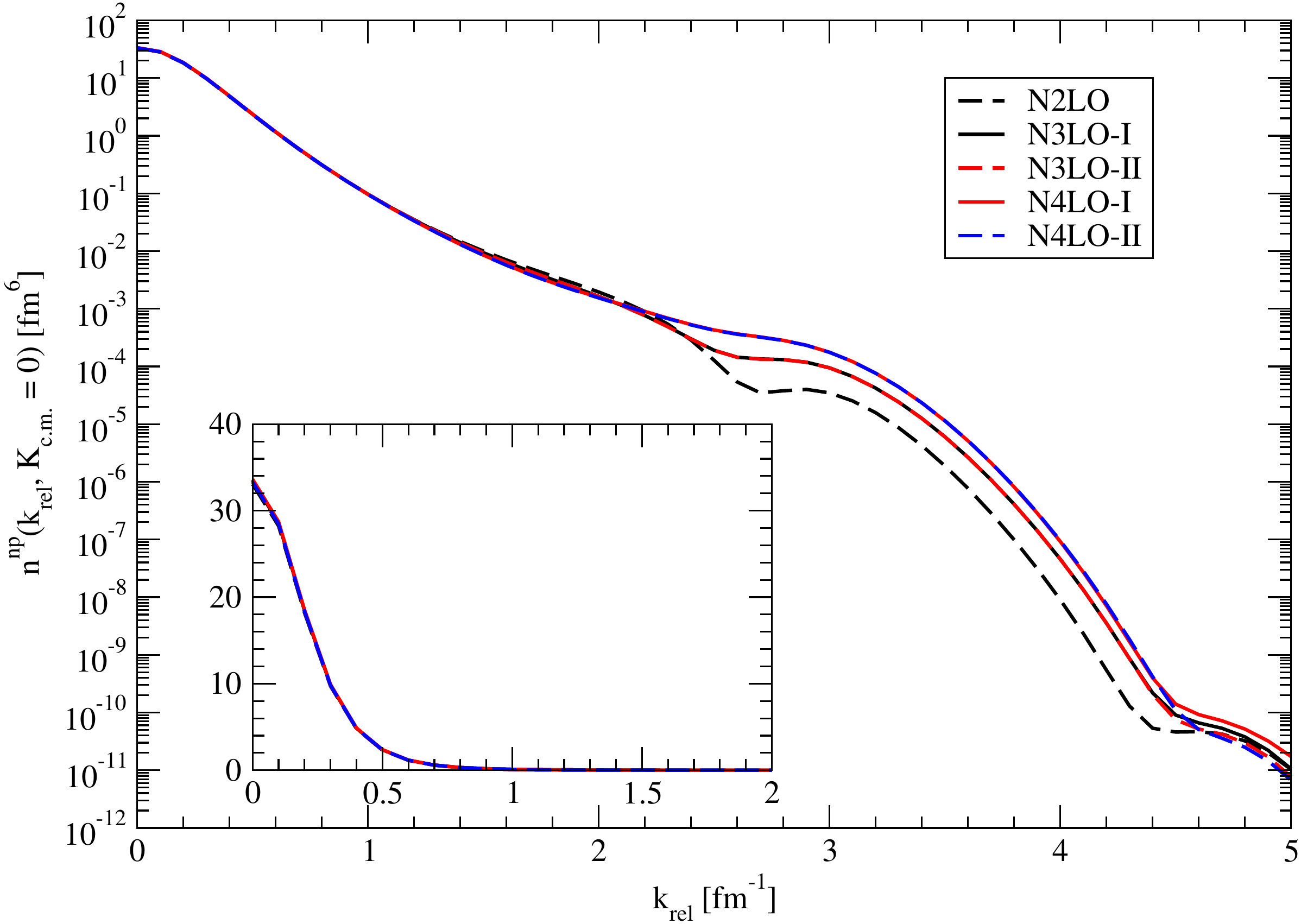}}
\caption{(Color online) The $np$ momentum
  distributions  $n^{np}(k_{rel},K_{c.m.}=0)$, calculated using
  only 2N (left panel) and 2N+3N (right panel) chiral interactions,
  with $\Lambda=500$ MeV.
  The different chiral orders are labelled as in the text.                 
  On the right panel, we indicate with ``I'' and ``II''
  the results
  obtained using the LECs of Table~\ref{tab:cdceI}
  and~\ref{tab:cdceII}, respectively.}
\label{fig:2bmd_np_bb_500}
\end{figure}

\begin{figure}[tbh!] 
\centering         
\scalebox{0.35}{\includegraphics{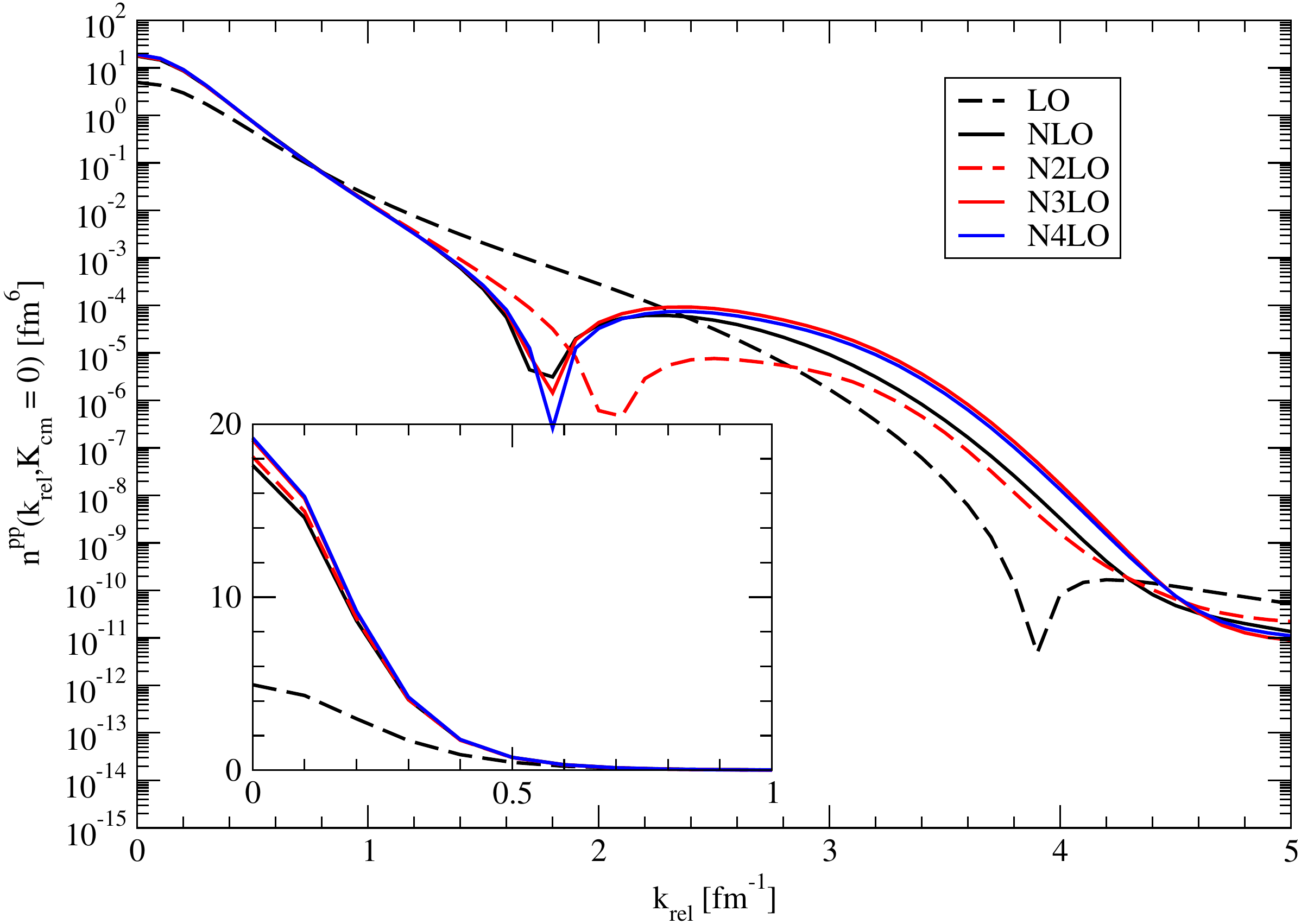}}
\scalebox{0.35}{\includegraphics{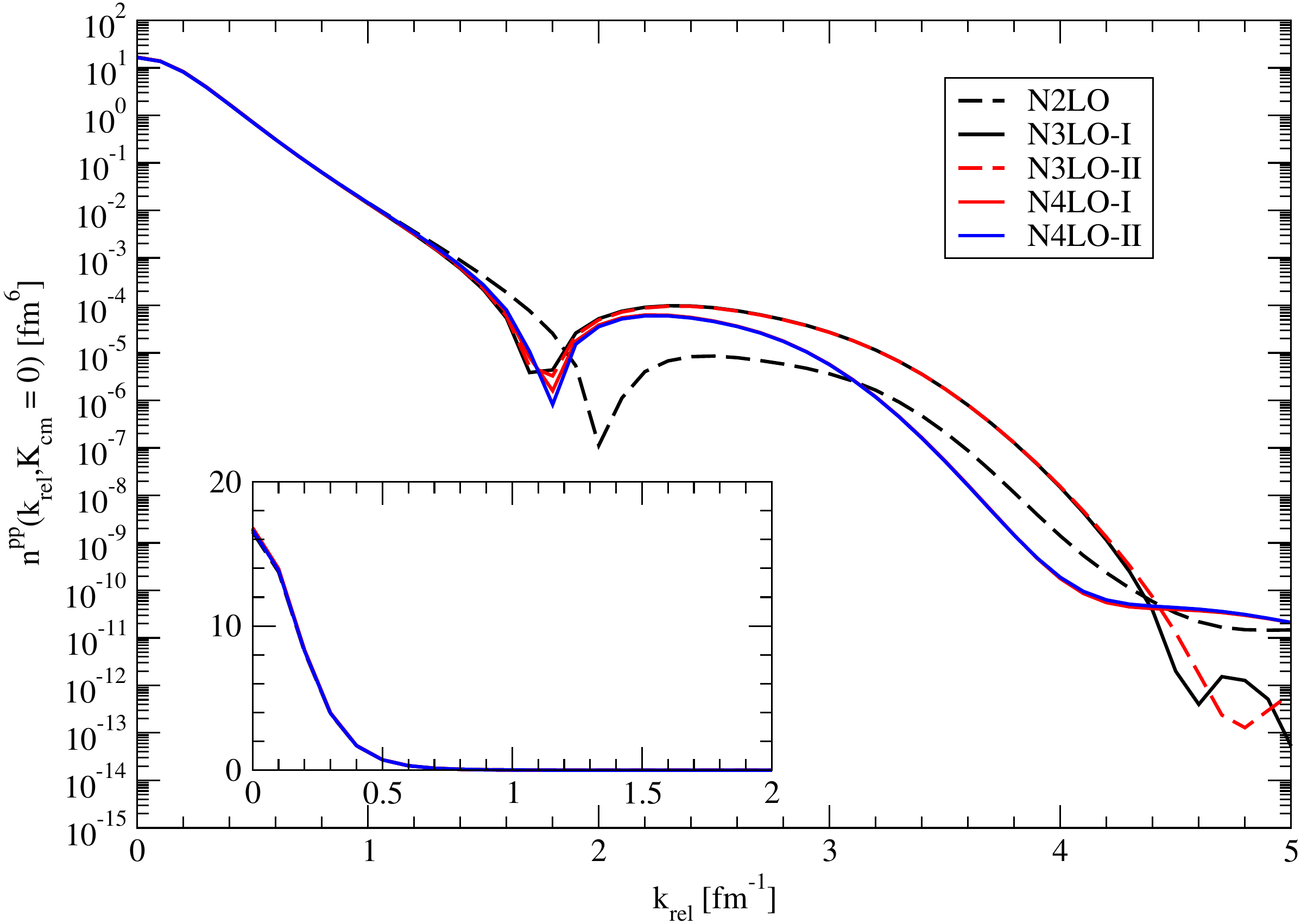}}
\caption{Same as Fig.~\ref{fig:2bmd_np_bb_500} but for the 
$pp$ pair.} 
\label{fig:2bmd_pp_bb_500}
\end{figure}

We now turn our attention to the 2N momentum distributions obtained
with the 2N chiral potentials without or with the 3N forces,
obtained as discussed in Sec.~\ref{subsec:th}. We begin with studying 
the order-by-order pattern, using the $\Lambda=500$ MeV cutoff
as an example. The results obtained with the other values
of $\Lambda$ display a similar behaviour. In
the left panel of Fig.~\ref{fig:2bmd_np_bb_500}
we show the BB $np$ momentum distribution $n^{np}(k_{rel},K_{c.m.}=0)$
obtained using only the 2N force at LO, NLO, N2LO, N3LO and N4LO.
In the right panel, we present the results for  $n^{np}(k_{rel},K_{c.m.}=0)$
including the 3N force, with LECs obtained from Table~\ref{tab:cdceI}
(model I) and~\ref{tab:cdceII} (model II), respectively. 
By inspection of the figures, we can conclude that the LO curve
has a 
the distinct behavior at small $k_{rel}$ compared with the other curves,
which suggests that the asymptotic part of the wave function at LO
is significantly different than at the higher orders. Furthermore, the
3N force contribution is very small, and therefore the difference
between the 3N force model is not visible. Finally, the N3LO and N4LO
curves are
very similar up to $k_{rel}\simeq 2.2$ fm$^{-1}$, indicating satisfactory 
order-by-order convergence at least in the region where
the distributions still have non-negligible size. In
Fig.~\ref{fig:2bmd_pp_bb_500} we show
the corresponding BB $pp$ momentum distributions. As we can see,
the same remarks apply in the $pp$ case as well. 

\begin{figure}[tbh!] 
\centering         
\scalebox{0.7}{\includegraphics{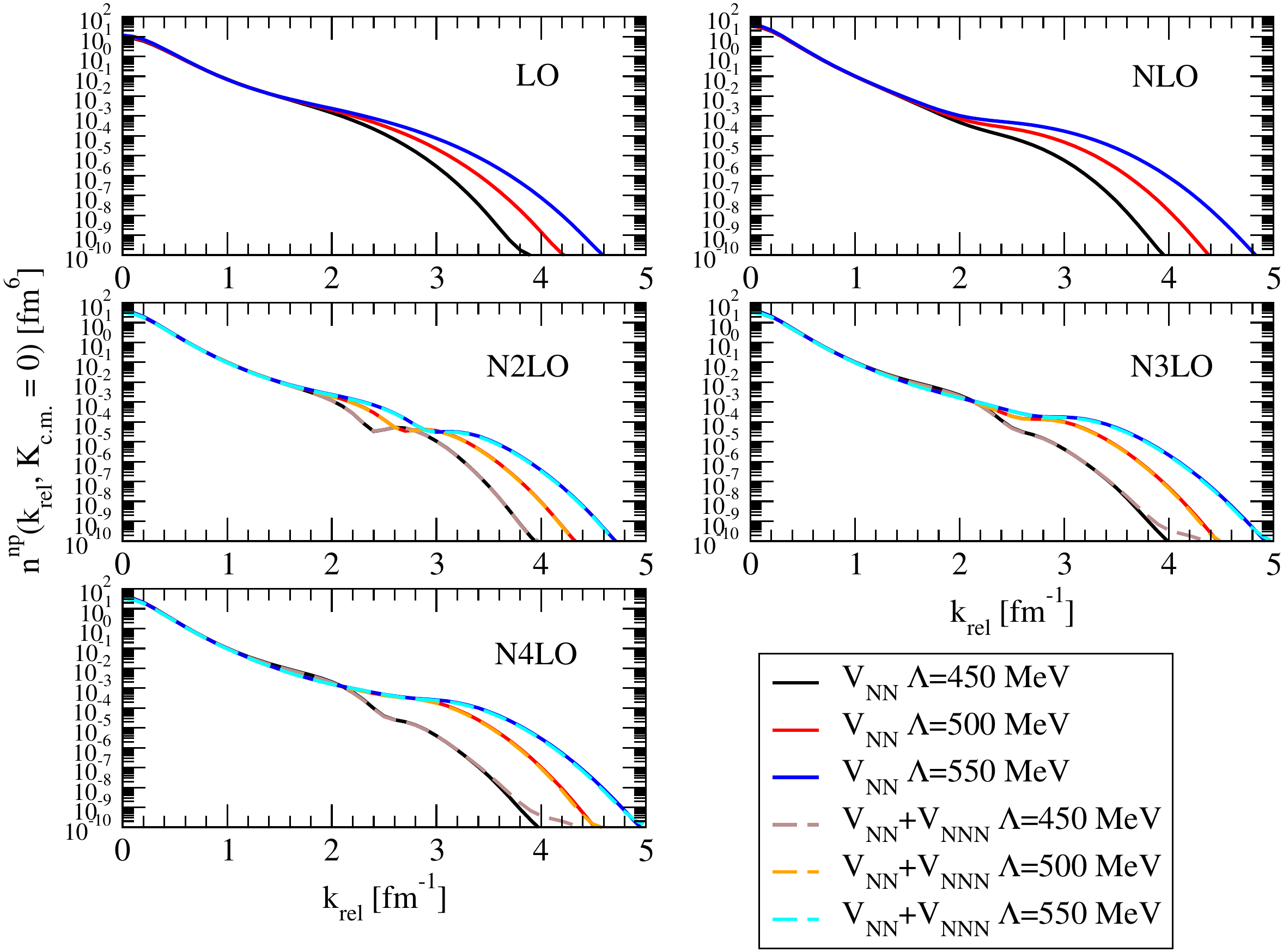}}
\caption{(Color online) The $np$ momentum distributions
  $n^{np}(k_{rel},K_{c.m.}=0)$, calculated using
  only 2N (solid lines) and 2N+3N (dashed lines) chiral interactions,
  at different chiral order and for three values of the  
  cutoff $\Lambda= 450, 500, 550$ MeV. The LECs of the
  3N interaction are those of Table~\ref{tab:cdceII}.} 
\label{fig:2bmd_bb_allL_np}
\end{figure}

\begin{figure}[tbh!] 
\centering         
\scalebox{0.7}{\includegraphics{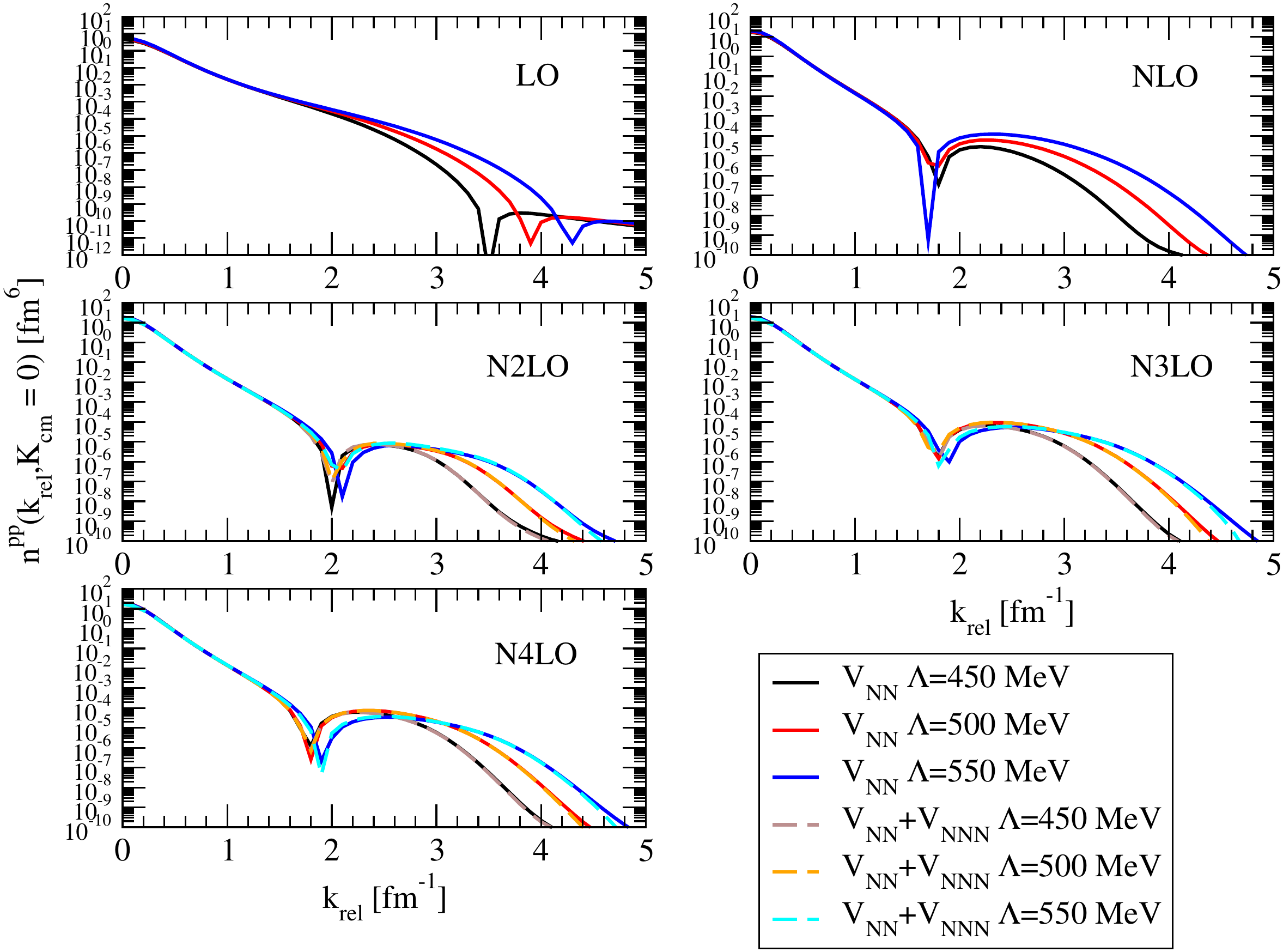}}
\caption{Same as Fig.~\ref{fig:2bmd_bb_allL_np} but for the $pp$ pair.} 
\label{fig:2bmd_bb_allL_pp}
\end{figure}

\begin{figure}[tbh!] 
\centering         
\scalebox{0.7}{\includegraphics{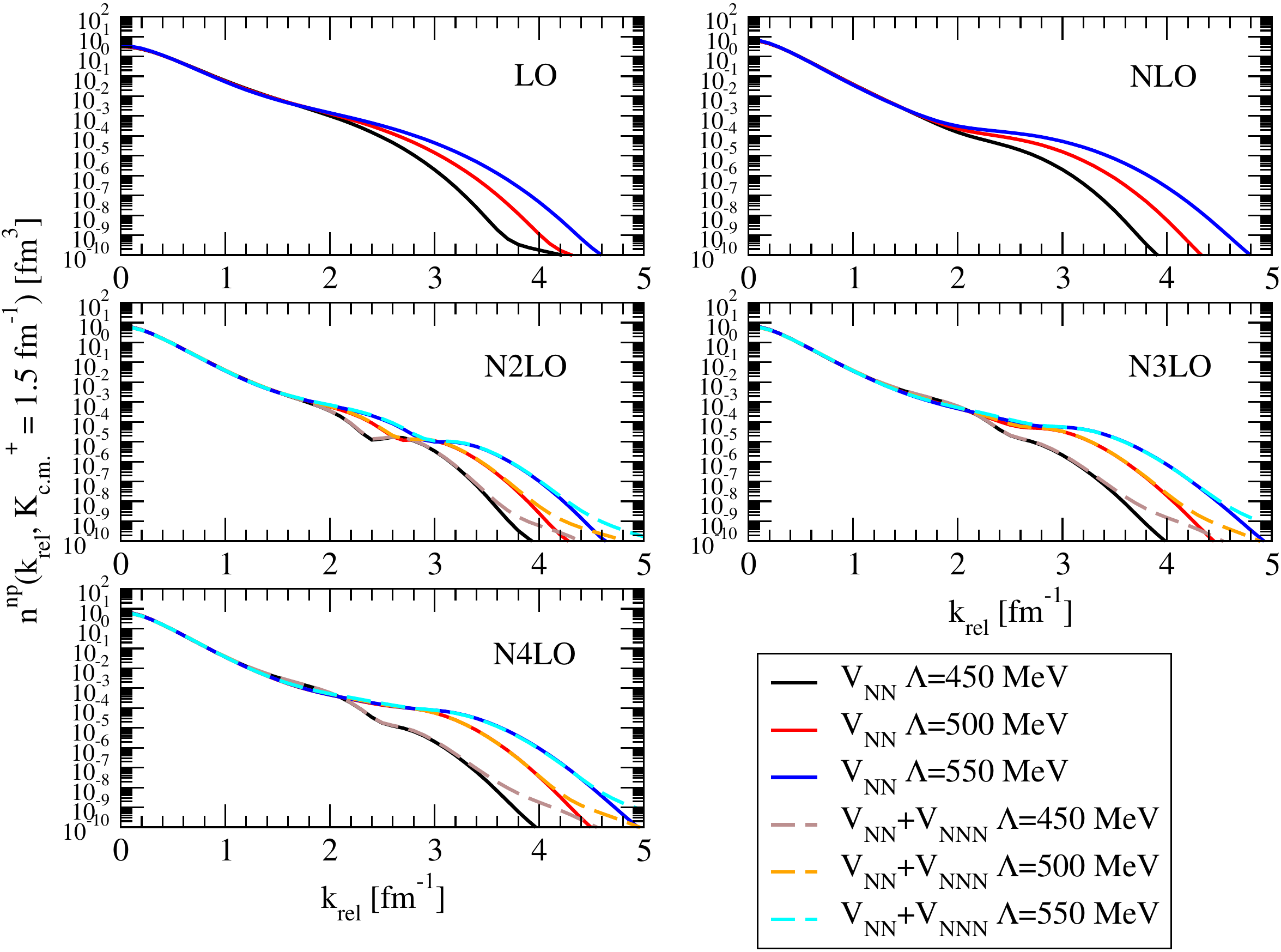}}
\caption{(Color online) The 2N momentum distributions
  $n^{np}(k_{rel},K_{c.m.}^+)$
  for $K_{c.m.}^+=+1.5$ fm$^{-1}$, calculated using
  only 2N (solid lines) and 2N+3N (dashed lines) chiral interactions,
  at different chiral order and for three values of the  
  cutoff $\Lambda= 450, 500, 550$ MeV. The LECs of the
  3N interaction are those of Table~\ref{tab:cdceII}.} 
\label{fig:2bmd_allL_kcm+1.5_np}
\end{figure}

\begin{figure}[tbh!] 
\centering         
\scalebox{0.7}{\includegraphics{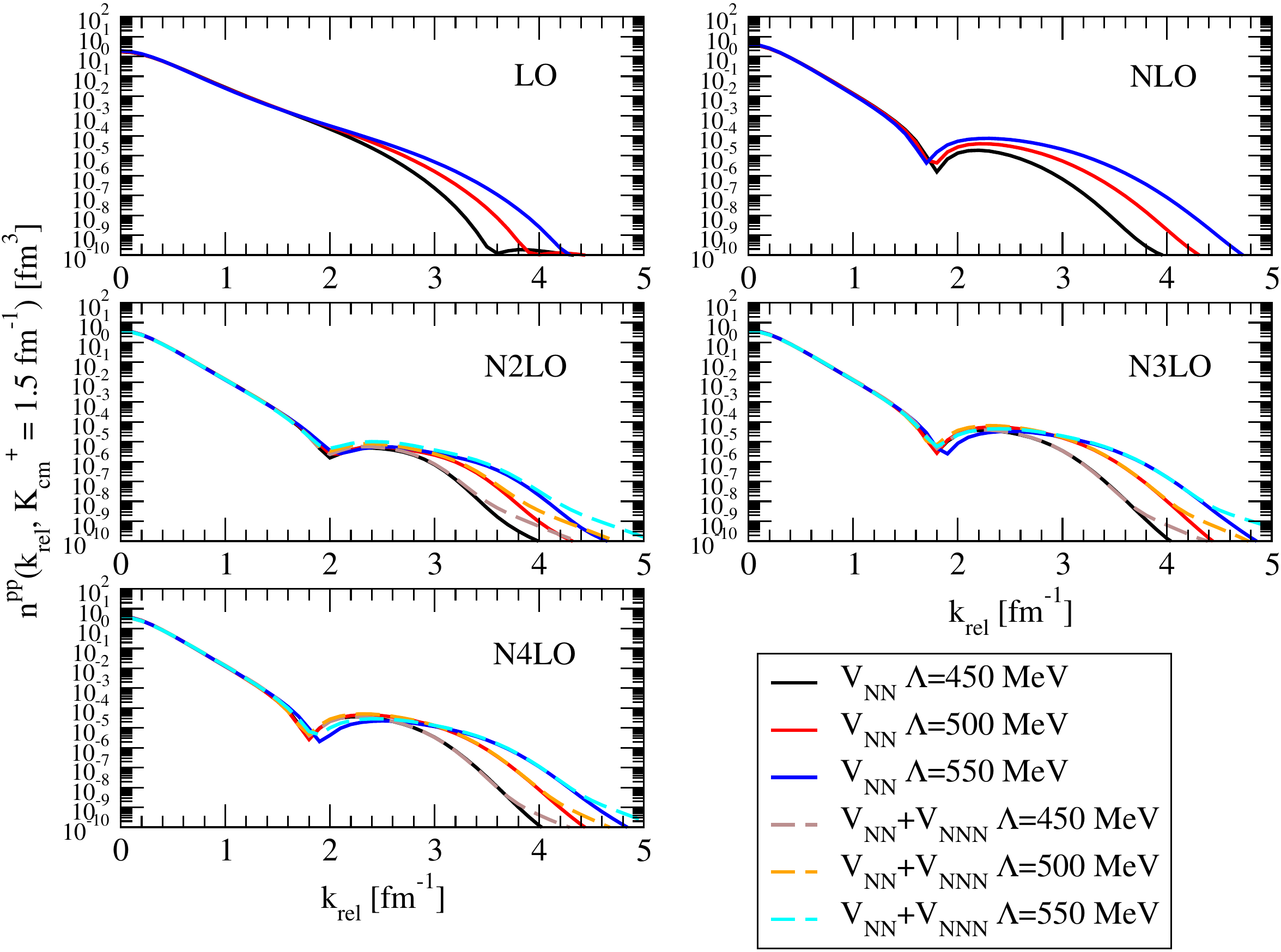}}
\caption{Same as Fig.~\ref{fig:2bmd_allL_kcm+1.5_np} but for the $pp$ pair.} 
\label{fig:2bmd_allL_kcm+1.5_pp}
\end{figure}

\begin{figure}[tbh!] 
\centering         
\scalebox{0.7}{\includegraphics{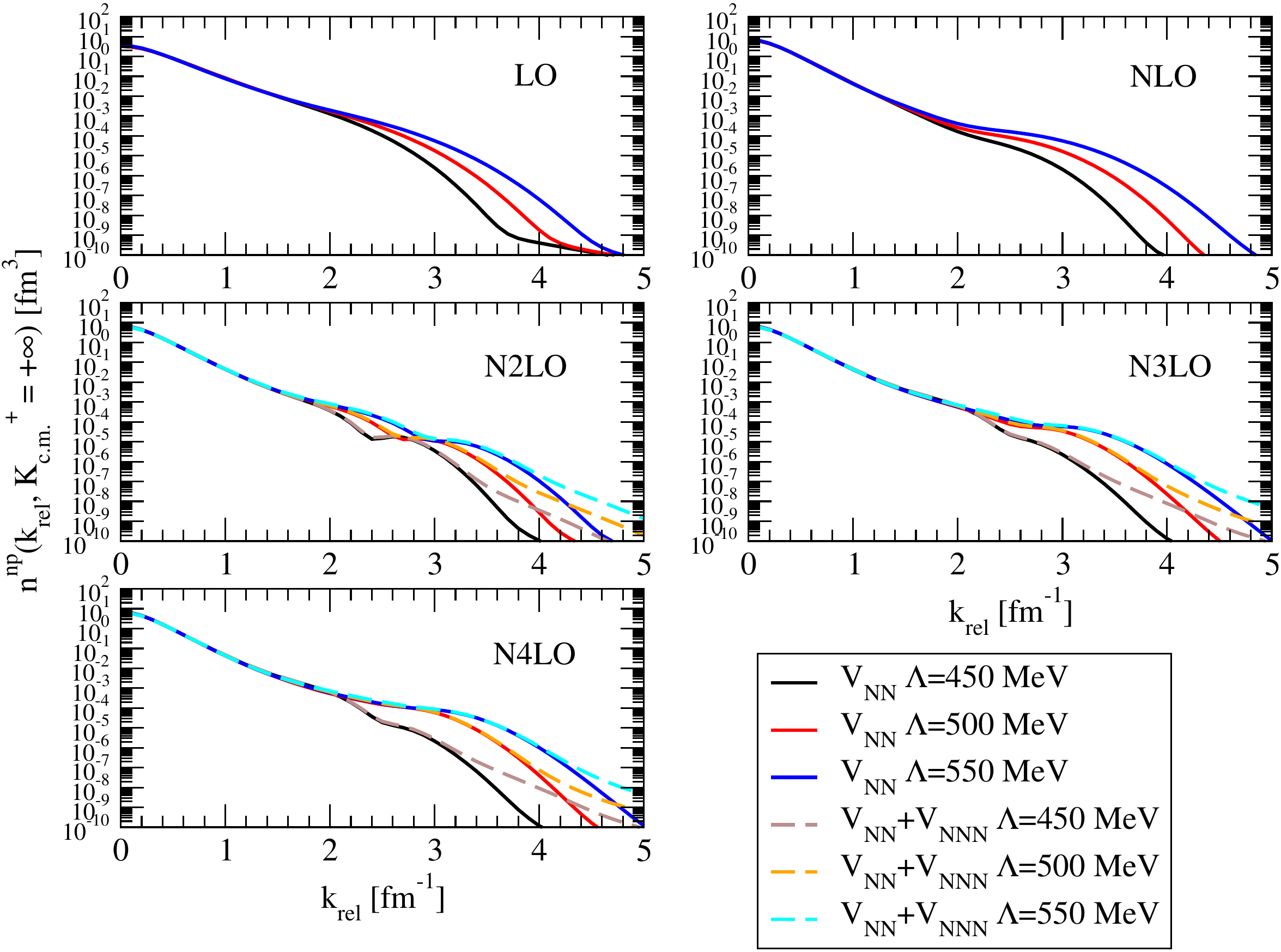}}
\caption{Same as Fig.~\ref{fig:2bmd_allL_kcm+1.5_np} but for
$K_{c.m.}^+=+\infty$.} 
\label{fig:2bmd_allL_kcm_infty_np}
\end{figure}

\begin{figure}[tbh!] 
\centering         
\scalebox{0.7}{\includegraphics{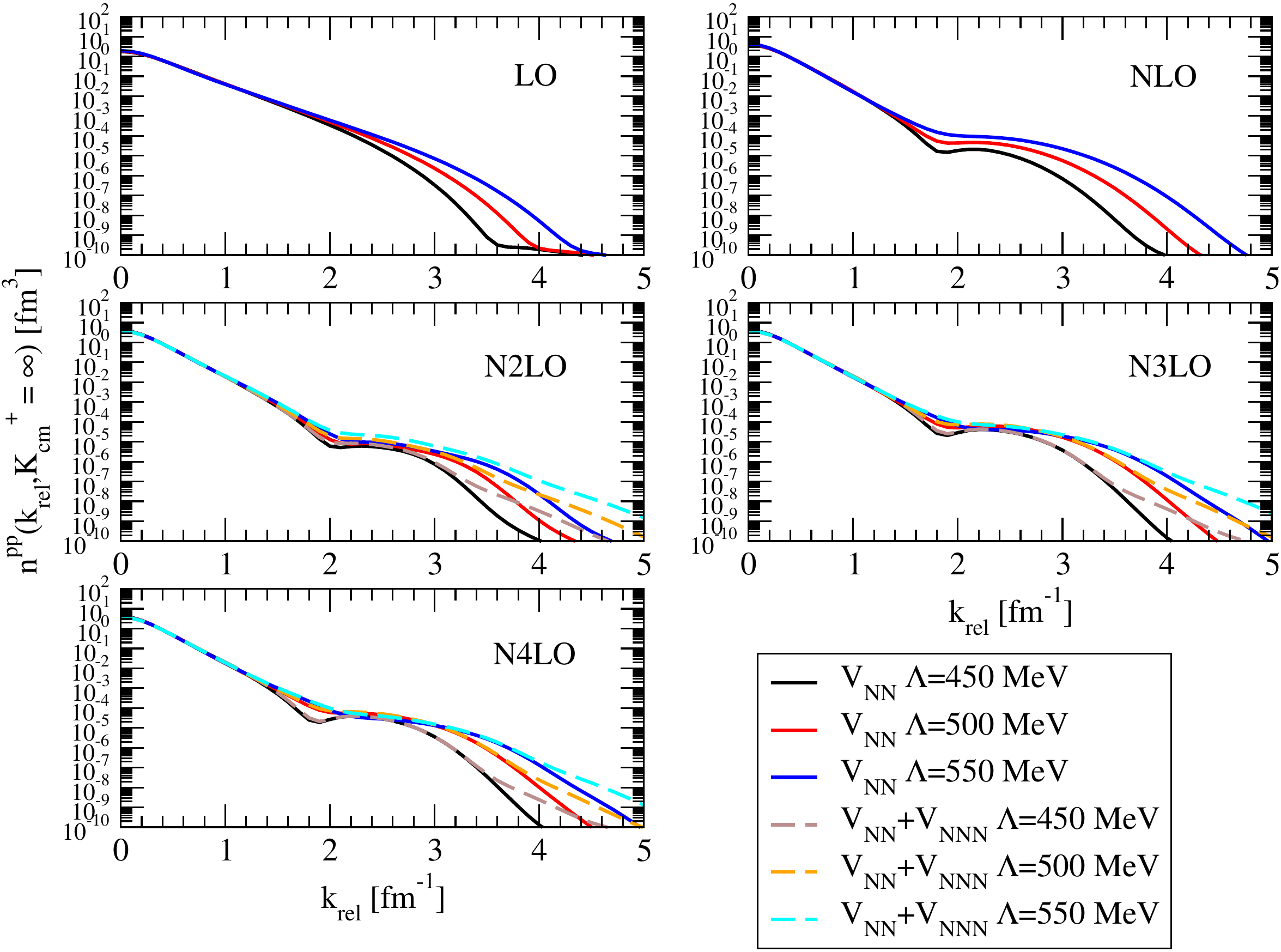}}
\caption{Same as Fig.~\ref{fig:2bmd_allL_kcm_infty_np} but for
the $pp$ pair.}
\label{fig:2bmd_allL_kcm_infty_pp}
\end{figure}

The BB 2N momentum distributions $n^{np}(k_{rel},K_{c.m.}=0)$ and
$n^{pp}(k_{rel},K_{c.m.}=0)$ calculated with and without 3N interaction,
at different chiral order and for different values of the
cutoff $\Lambda$, are shown in Figs.~\ref{fig:2bmd_bb_allL_np}
and~\ref{fig:2bmd_bb_allL_pp}, respectively.
The results for the
2N momentum distributions $n^{np/pp}(k_{rel},K_{c.m.}^+)$
for $K_{c.m.}^+=+1.5$ fm$^{-1}$ and $K_{c.m.}^+=+\infty$ are
shown in
Figs.~\ref{fig:2bmd_allL_kcm+1.5_np},~\ref{fig:2bmd_allL_kcm+1.5_pp},~\ref{fig:2bmd_allL_kcm_infty_np}
and~\ref{fig:2bmd_allL_kcm_infty_pp}.                                  
By inspection of all the figures we can conclude that
we have essentially no cutoff dependence below 
$k_{rel} \simeq 2.2-2.5$
fm$^{-1}$, and increasingly strong cutoff dependence above it. Furthermore, the 3N force
contributions are visible only for $k_{rel}\ge 3.0-3.5$ fm$^{-1}$.
Note, however, that above 
$k_{rel} \simeq 2.5$ fm$^{-1}$ all momentum distributions are so small that the differences are
of no practical relevance, see next. 

Calculating the integrated SRCs as defined in
Eqs.~(\ref{eq:n-bb})--(\ref{eq:n}), we obtain the values displayed in
Tables~\ref{tab:nnn_chir_np} and~\ref{tab:nnn_chir_pp}
for $np$ and $pp$ SRCs, respectively.                                 
For these ``observables", as well, we find that   
order-by-order convergence is satisfactory and cutoff dependence is weak.  
This implies that the contributions from the region
$k_{rel}\ge 2.2$ fm$^{-1}$ are essentially negligible.       
\begin{table}[t]
  \caption{\label{tab:nnn_chir_np}The integrated $np$
  SRC-probabilities, as defined in Eqs.~(\ref{eq:n-bb})--(\ref{eq:n}),
  obtained with chiral potentials, at different chiral orders,
  for different values of the cutoff $\Lambda$, 450, 500 and 550 MeV,
  also with the inclusion of the 3N force (lines labelled
  N2LO/N2LO at N2LO, N3LO/N3LO-I and N3LO/N3LO-I at N3LO,
  N4LO/N4LO-I and N4LO/N4LO-I at N4LO). The labels ``I'' and ``II''
  refer to the LECs of Table~\ref{tab:cdceI} or~\ref{tab:cdceII},
  respectively.}
  \begin{tabular}{c|ccc|ccc|ccc|ccc}
    \hline
    \hline
    &\multicolumn{3}{|c|}{$N^{BB}$} & \multicolumn{3}{|c|}{$N^{SRC,BB}$}
    &\multicolumn{3}{|c|}{$N^{SRC}(k_{rel}^-)$}
    &\multicolumn{3}{|c}{$N$}\\ 
\hline
Model/$\Lambda$ [MeV] & 450 & 500 & 550 &
450 & 500 & 550 & 450 & 500 & 550 & 450 & 500 & 550 \\
\hline
LO           & 2.731 & 2.829 & 3.051
             & 0.094 & 0.120 & 0.144 & 0.089 & 0.112 & 0.126
             & 1.999 & 1.999 & 1.998 \\
NLO          & 5.896 & 6.054 & 6.458
             & 0.047 & 0.066 & 0.096 & 0.016 & 0.024 & 0.037
             & 1.998 & 1.998 & 1.997 \\
\hline 
N2LO         & 5.977 & 6.127 & 6.236
             & 0.087 & 0.118 & 0.141 & 0.029 & 0.038 & 0.045
             & 1.998 & 1.997 & 1.997 \\
N2LO/N2LO    & 5.844 & 5.831 & 5.827
             & 0.086 & 0.114 & 0.135 & 0.030 & 0.040 & 0.050
             & 1.998 & 1.998 & 1.998 \\
\hline
N3LO         & 6.443 & 6.314 & 6.317
             & 0.131 & 0.112 & 0.122 & 0.039 & 0.039 & 0.044
             & 1.997 & 1.997 & 1.997 \\
N3LO/N3LO-I  & 5.823 & 5.884 & 5.907
             & 0.121 & 0.107 & 0.117 & 0.042 & 0.043 & 0.051
             & 1.998 & 1.998 & 1.998 \\
N3LO/N3LO-II & 5.817 & 5.865 & 5.890
             & 0.121 & 0.107 & 0.117 & 0.042 & 0.043 & 0.050
             & 1.998 & 1.998 & 1.998 \\
\hline
N4LO         & 6.360 & 6.345 & 6.266
             & 0.125 & 0.119 & 0.129 & 0.038 & 0.042 & 0.047
             & 1.997 & 1.997 & 1.998 \\
N4LO/N4LO-I  & 5.823 & 5.911 & 5.915
             & 0.116 & 0.114 & 0.125 & 0.041 & 0.047 & 0.054
             & 1.998 & 1.998 & 1.998 \\
N4LO/N4LO-II & 5.809 & 5.868 & 5.857
             & 0.116 & 0.113 & 0.123 & 0.040 & 0.045 & 0.051
             & 1.998 & 1.998 & 1.998 \\
\hline
  \end{tabular}
\end{table}

\begin{table}[t]
  \caption{\label{tab:nnn_chir_pp} Same as
    Table~\ref{tab:nnn_chir_np} but for the $pp$ pair.}
  \begin{tabular}{c|ccc|ccc|ccc|ccc}
    \hline
    \hline
    &\multicolumn{3}{|c|}{$N^{BB}$} & \multicolumn{3}{|c|}{$N^{SRC,BB}$}
    &\multicolumn{3}{|c|}{$N^{SRC}(k_{rel}^-)$}
    &\multicolumn{3}{|c}{$N$}\\ 
\hline
Model/$\Lambda$ [MeV] & 450 & 500 & 550 &
450 & 500 & 550 & 450 & 500 & 550 & 450 & 500 & 550 \\
\hline
LO           & 1.049 & 1.087 & 1.165 & 0.016 & 0.020 & 0.023
& 0.030 & 0.040 & 0.045
& 0.999 & 0.999 & 0.999 \\
NLO          & 1.976 & 2.014 & 2.109 & 0.002 & 0.004 & 0.009
& 0.002 & 0.004 & 0.010
& 0.999 & 0.999 & 0.998 \\
\hline 
N2LO         & 1.975 & 2.009 & 2.036 & 0.002 & 0.002 & 0.003
& 0.003 & 0.004 & 0.005
& 0.998 & 0.998 & 0.998 \\
N2LO/N2LO    & 1.943 & 1.935 & 1.932 & 0.002 & 0.002 & 0.002
& 0.003 & 0.004 & 0.006
& 0.999 & 0.998 & 0.998 \\
\hline
N3LO         & 2.083 & 2.061 & 2.060 & 0.004 & 0.007 & 0.006
& 0.003 & 0.007 & 0.009
& 0.998 & 0.998 & 0.998 \\
N3LO/N3LO-I  & 1.928 & 1.952 & 1.958 & 0.004 & 0.007 & 0.006
& 0.004 & 0.009 & 0.011
& 0.998 & 0.999 & 0.999 \\
N3LO/N3LO-II & 1.927 & 1.948 & 1.953 & 0.004 & 0.007 & 0.006
& 0.004 & 0.008 & 0.011
& 0.998 & 0.999 & 0.999 \\
\hline
N4LO         & 2.064 & 2.070 & 2.048 & 0.004 & 0.005 & 0.004
& 0.004 & 0.008 & 0.009
& 0.998 & 0.998 & 0.998 \\
N4LO/N4LO-I  & 1.929 & 1.960 & 1.962 & 0.004 & 0.006 & 0.004
& 0.004 & 0.009 & 0.012
& 0.998 & 0.999 & 0.999 \\
N4LO/N4LO-II & 1.926 & 1.949 & 1.945 & 0.004 & 0.005 & 0.004
& 0.004 & 0.009 & 0.010
& 0.998 & 0.999 & 0.999 \\
\hline
  \end{tabular}
\end{table}

Earlier in the paper, we noted that the momentum distributions we calculated with 
chiral interactions die out at a faster rate than those obtained with phenomenological potentials.
In Sect.~\ref{subsec:single-n}, we pointed out that such feature may be expected given the softer nature of     
chiral forces. While this is a correct observation 
within the spectrum of interactions considered here, it is 
important to note that the chiral nature of an interaction does not 
necessarily bring additional softness. To support this
statement, we refer to Ref.~\cite{loc_chi}, where predictions for 1N and 2N
momentum distributions in $A \le $16 are shown. In that work it is concluded that,
when {\it local} chiral interactions are 
employed, the resulting momentum distributions are {\it consistent} with those obtained from local phenomenological 
potentials. In fact, the local 2N chiral interactions (at N2LO) applied in Ref.~\cite{loc_chi} and developed
in Refs.~\cite{Gez+13,Gez+14} predict a $D$-state probability for the deuteron ranging between 5.5 and 
6.1\%, values which are typical for the ``hardest" local potentials. 

Therefore, once again, the local {\it vs.} non-local nature of the 2N force (by far the largest contribution 
to the 1N and 2N momentum distributions, as
we have observed on several occasions), is a major factor in determining the characteristics of momentum 
distributions in nuclei and, particularly, their short-range part.

\section{Conclusions and outlook}
\label{sec:concl} 

We have presented predictions for 1N and 2N momentum distributions in the 
deuteron and in $^3$He.                           
We have employed state-of-the-art chiral 2N potentials (with or without the 
leading chiral 3N force) and also, for the purpose
of comparison and 
validation of our tools, older potentials plus 3N force, either fully phenomenological or based on meson theory.          
A main motivation was to explore the short-range few-nucleon dynamics as predicted by these diverse 
interactions.  One of our 
findings is that, regardless the 2N force model, the contribution from 3N
forces is always very weak. 

We have also quantified and pointed out, 
as appropriate, any significant model dependence of SRCs. We noted that SRCs in the lightest few-nucleon systems
may impact the extraction of (semi-)empirical information for heavier nuclei, all the way to      
extrapolations in nuclear matter. In fact, 
for heavier systems, integrated SRC probabilities 
have been extracted from empirical information together with some model-dependent assumptions~\cite{CLAS,CLAS2}.

Although potentials based on chiral EFT may be expected to produce weaker SRC than purely phenomenological 
or meson-exchange ones, 
on several occasions the discussion of model dependence led back to considerations of locality {\it vs.} 
non-locality of the underlying 2N force, rather than ``chiral" {\it vs.} ``non-chiral".
We find this to be an important issue, extensively debated in the literature 
of the 1990's~\cite{MSS96,Polls98,MP99} and now re-emerging 
in the light of new stimulating discussions. 

The 2N potentials considered here have an established success record with low-energy predictions, 
such as the structure of light and medium-mass nuclei as well as the 
properties of nuclear matter. But, as shown above, they differ
considerably in their high-momentum components.
Thus, one may raise the question of     
whether the predictions of these 
potentials in the high-momentum regions probed by the JLab measurements are reliable or, to a
certain extent, arbitrary. Note that                                              
there is no physical reason why the off-shell behavior of, say, AV18, should be preferable as compared
to other potentials. In fact, on fundamental grounds off-shell behavior is not observable and, therefore, 
``empirical" information related to the (as we have shown, very diverse) off-shell nature of the potentials 
is highly model dependent. 
To stress this point even more: 2N potentials which are known as $V_{low-k}$~\cite{BKS03} are typically cut
off between 1.5 and 2 fm$^{-1}$ and, thus, produce essentially zero SRC. However, these low-momentum interactions are
highly successful in {\it ab initio} nuclear structure calculations and, thus, are valid 2N potentials. 

Considering all of the above, 
our investigation prompts us to  conclude the following: since 
SRC ``empirical" information is in part based on the high-momentum behavior of
a specific 2N interaction~\cite{CLAS,CLAS2}, 
the reported information should be understood as carrying very large uncertainty, and such uncertainty should be
quantified and stated. 

In the near future, we plan to extend our study to triton and $^4$He.
In both the $A=3$ nuclei, we will look 
into other interesting quantities
which are related to momentum distributions and SRC, such as the ratio 
of $pp$ to $pn$ 2N momentum distributions {\it vs.} $k_{rel}$, and the
ration of $pp$ to $np$ pairs in the back-to-back case. 
For $^4$He, the ratio of $pp$ to $np$ pairs is also of interest,
as information on this ratio as extracted from high momentum-transfer
triple-coincidence 
measurements of $^4$He($e,e'\;p\;N$) is reported in
Ref.~\cite{Korover}. Similar measurements are present in 
Ref.~\cite{Sub+2008} for $^{27}$Al, and in 
Ref.~\cite{Hen+2014} for $^{208}$Pb.

\section*{Acknowledgments}
The work of F.S. and R.M. was supported by 
the U.S. Department of Energy, Office of Science, Office of Basic Energy
Sciences, under Award Number DE-FG02-03ER41270. 
Computational resources provided by the INFN-Pisa Computer Center
are gratefully acknowledged.


\begin{thebibliography}{100}

\bibitem{CLAS} K.S. Egiyan {\it et al.}, Phys. Rev. Lett. {\bf 96}, 082501 (2006), and references therein.
\bibitem{CLAS2} K.S. Egiyan {\it et al.}, CLAS-NOTE 2005-004, 2005, www1.jlab.org/ul/Physics/Hall-B/clas.
\bibitem{CLAS3} K.S. Egiyan {et al.}, Phys. Rev. C {\bf 68}, 014313 (2003). 
\bibitem{src} Michael McGauley and Misak M. Sargsian, arXiv:1102.3973v3 (2012). 
\bibitem{Pia+} E. Piasetzky {\it et al.}, Phys. Rev. Lett. {\bf 97}, 162504 (2006).
\bibitem{Egi06} K.S. Egiyan {\it et al.}, Phys. Rev. Lett. {\bf 96}, 082501 (2006).
\bibitem{FS14} F. Sammarruca, Phys. Rev. C {\bf 90 }, 064312 (2014); and references therein. 

\bibitem{Tang} A. Tang {\it et al.}, Phys. Rev. Lett. {\bf 90}, 042301 (2003). 
\bibitem{Korover} I. Korover {\it et al.} (Jefferson Lab Hall A Collaboration), Phys. Rev. Lett. {\bf 113}, 022501 
(2014). 
\bibitem{Shneor} R. Shneor {\it et al.} (Jefferson Lab Hall A Collaboration), Phys. Rev. Lett. {\bf 99}, 072501 (2007).
\bibitem{Subedi} R. Subedi {\it et al.}, Science {\bf 320}, 1476 (2008).
\bibitem{Baghda} H. Baghdasaryan {\it et al.} (CLAS Collaboration), Phys. Rev. Lett. {\bf 105}, 222501 (2010). 
\bibitem{src2015} F. Sammarruca, Phys. Rev. C {\bf 92}, 044003 (2015).

\bibitem{av18} R.B. Wiringa, V.G.J. Stoks, and R. Schiavilla, Phys. Rev. C {\bf 51}, 38 (1995).              
\bibitem{Nij} V.G.J. Stoks, R.A.M. Klomp, C.P.F. Terheggen, and J.J. de Swart, Phys. Rev. C {\bf 49}, 2950 (1994). 
\bibitem{CD} R. Machleidt, Phys. Rev. C {\bf 63}, 024001 (2001). 
\bibitem{EM03} D.R. Entem and R. Machleidt, Phys. Rev. C {\bf 68}, 041001, (2003). 
\bibitem{chinn5} E. Marji, A. Canul, Q. MacPherson, R. Winzer, Ch. Zeoli, D.R. Entem, and R. Machleidt,
Phys. Rev. C {\bf 88}, 054002 (2013).
\bibitem{ME11} R. Machleidt and D.R. Entem,  Phys. Rep. {\bf 503}, 1 (2011).
\bibitem{MSS96} R. Machleidt, F. Sammarruca, and Y. Song, Phys. Rev. C {\bf 53}, 1483 (1996). 
\bibitem{Polls98} A. Polls, H. M{\"u}ther, R. Machleidt, and M. Hjorth-Jensen, Phys. Lett. B {\bf 432}, 1 (1998).
\bibitem{MP99} H. M{\"u}ther and A.Polls, Phys. Rev. C {\bf 61}, 014304 (1999). 
\bibitem{EMN} D.R. Entem, R. Machleidt, and Y. Nosyk, Phys. Rev. C {\bf 96}, 024004 (2017).

\bibitem{Hofe+} M. Hoferichter, J. Ruiz, de Elvira, B. Kubis, and U.-G. Meissner, Phys. Rev. Lett. {\bf 115},
192301 (2015); Phys. Rep. {\bf 625}, 1 (2016). 
\bibitem{UIX} B.S. Pudliner, V.R. Pandharipande, J. Carlson, and 
R.B. Wiringa, Phys. Rev. Lett. {\bf 74}, 4396 (1995).     
\bibitem{TM} S.A. Coon and H.K. Han, Few-Body Systems {\bf 30}, 131 (2001).       

\bibitem{Alv13} M. Alvioli, C. Ciofi degli Atti, L.P. Kaptari, C.B. Mezzetti, and H. Morita,
Phys. Rev. C {\bf 87}, 034603 (2013).
\bibitem{Alv16} M. Alvioli, C. Ciofi degli Atti, and H. Morita, 
Phys. Rev. C {\bf 94}, 044309 (2016).
\bibitem{Wir_web}R.B.\ Wiringa, R.\ Schiavilla, S.\ Pieper, and J.\ Carlson,
  Phys.\ Rev.\ C {\bf 89}, 024305 (2014); R.B.\ Wiringa,
  {\tt https://www.phy.anl.gov/theory/research/momenta2/}

\bibitem{Viv06} M.\ Viviani {\it et al.}, Few-Body Syst. {\bf 39}, 159 (2006)
\bibitem {Kie08}A.\ Kievsky {\it et al.}, J.\ Phys. G {\bf 35}, 063101 (2008)
\bibitem{Mar12} L.E.\ Marcucci, A.\ Kievsky, S.\ Rosati, R.\ Schiavilla, and
M.\ Viviani, Phys.\ Rev.\ Lett.\ {\bf 108}, 052502, (2012); Erratum:
Phys.\ Rev.\ Lett.\ {\bf 121}, 049901 (2018).
\bibitem{nuclmatt18} F.\ Sammarruca, L.E.\ Marcucci, L.\ Coraggio,
  J.W.\ Holt, N.\ Itaco, and R.\ Machleidt, arXiv:1807.06640.     
\bibitem{Kre12} H.\ Krebs, A.\ Gasparyan, and E.\ Epelbaum, Phys.\ Rev.\
C {\bf 85}, 054006 (2012).

\bibitem{VDC} J.G. Van der Corput, Akademie van Wetenschappen {\bf 38}, 813 (1935).
\bibitem{Frank93} L. L. Frankfurt, M.I. Strikman, D.B. Day, and M. Sargsyan, Phys. Rev. C {\bf 48}, 2451 (1993).
\bibitem{Ciof} C. Ciofi degli Atti and S. Simula, Phys. Rev. C {\bf 53}, 1689 (1996).
\bibitem{Nogga} A. Nogga {\it et al.}, Phys. Rev. C {\bf 67}, 034004 (2003).
\bibitem{3nf1} V. Bernard, E. Epelbaum, H. Krebs, and Ulf-G. Meissner, Phys. Rev. C {\bf 77}, 064004 (2008);
Phys. Rev. C {\bf 84}, 054001 (2011). 
\bibitem{3nf2} H. Krebs, A. Gasparyan, and E. Epelbaum, Phys. Rev. C {\bf 85}, 054006 (2012); Phys. Rev. C
{\bf 87}, 054007 (2013). 
\bibitem{3nf3} L. Girlanda, A. Kievsky, and M. Viviani, Phys. Rev. C {\bf 84}, 014001 (2011).                    
\bibitem{Pia13} E. Piasetzky, O. Hen, and L.B. Weinstein, AIP Conf. Proc. 1560, 355 (2013). 
\bibitem{loc_chi} D. Lonardoni, S. Gandolfi, X.B. Wang, and J. Carlson, arXiv:1804.08027.        
\bibitem{Gez+13} A. Gezerlis, I. Tews, E. Epelbaum, S. Gandolfi, K. Hebeler, A. Nogga, and A. Schwenk, Phys. Rev. Lett {\bf 111}, 032501 (2013).                    
\bibitem{Gez+14} A. Gezerlis, I. Tews, E. Epelbaum, M. Freunek, S. Gandolfi, K. Hebeler, A. Nogga, and A. Schwenk, Phys. Rev. C {\bf 90}, 054323 (2014).                    
\bibitem{BKS03} S.K. Bogner, T.T.S. Kuo, and A. Schwenk, Phys. Rep. {\bf 386}, 1 (2003).       
\bibitem{Sub+2008} R. Subedi {\it et al.}, Science {\bf 320}, 1476 (2008).                                  
\bibitem{Hen+2014} O. Hen {\it et al.}, Science {\bf 346}, 614 (2014).                                  

\end{thebibliography}
\end{document}